\documentclass{appolb}
\usepackage{epsfig,psfig,cite}

\def\be{\begin{equation}}
\def\ee{\end{equation}}
\def\bea{\begin{eqnarray}}
\def\eea{\end{eqnarray}}
\def\pom{I\!\!P}
\def\pro{\mbox{\scriptsize p}}
\def\max{\mbox{\scriptsize max}}
\def\sca{\mbox{\scriptsize scalar}}

\def\aem{\alpha_{\mbox{\scriptsize em}}}
\def\xbj{x_{\mbox{\scriptsize Bj}}}

\def\x{x_\perp}
\def\y{y_\perp}
\def\sc{\scriptsize}

\newcommand{\ks}{k\!\!\!/}

\newcommand{\ppl}{p\!\!/}
\newcommand{\epsilons}{\epsilon\!\!/}

\begin{document}
\pagestyle{plain}
\eqsec  
\title{Diffractive Electroproduction
\thanks{lectures at the Cracow School of Theoretical Physics, XXXIX Course,
Zakopane 1999}%
}
\author{A. Hebecker
\address{Institut f\"ur Theoretische Physik der Universit\"at Heidelberg\\
Philosophenweg 16, D-69120 Heidelberg, Germany}
}
\maketitle
\begin{abstract}
In these lectures, a simple introduction to the phenomenon of diffraction 
in deep inelastic scattering and its theoretical description is given. 
While the main focus is on the diffractive structure function $F_2^D$, some 
issues in diffractive vector meson production are also discussed. 
\end{abstract}
  
\section{Introduction}
In the following, the term `diffractive electroproduction' or `diffractive 
deep inelastic scattering (DIS)' is used to characterize those processes in 
small-$x$ DIS where, in spite of the high virtuality of the exchanged 
photon, the proton target remains intact or almost intact. Thus, the 
term is used synonymously with the expression `rapidity gap events in 
DIS'. 

\begin{figure}[ht]
\begin{center}
\vspace*{.2cm}
\parbox[b]{9cm}{\psfig{width=9cm,file=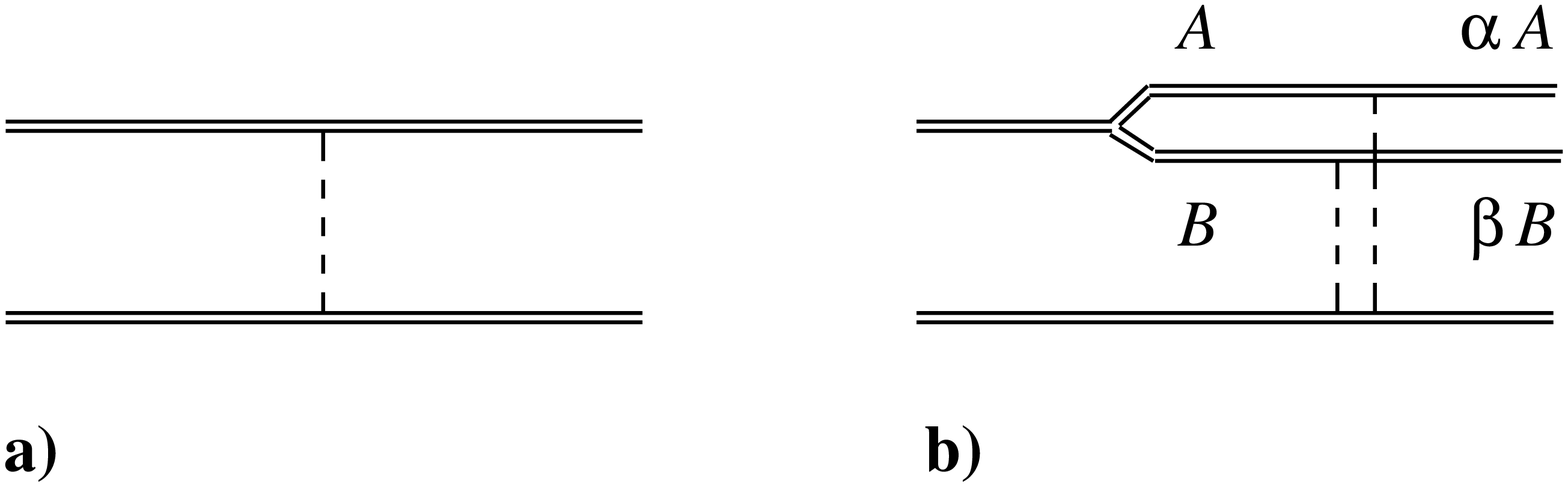}}\\
\end{center}
\refstepcounter{figure}
\label{fig:gwp}
{\bf Figure \ref{fig:gwp}:} Elastic scattering (a) and diffractive 
dissociation (b) in the physical picture of Good and Walker~\cite{gw}. 
\end{figure}

Historically, the term diffraction is derived from optics, where it 
describes the deflection of a beam of light and its decomposition into 
components with different frequencies. In high energy physics, it was 
originally used for small-angle elastic scattering of hadrons 
(Fig.~\ref{fig:gwp}a). If one of the hadrons, say the projectile, is 
transformed into a set of two or more final state particles, the process 
is called diffractive dissociation or inelastic diffraction. Good and 
Walker have pointed out that a particularly intuitive physical picture of 
such processes emerges if the projectile is described as a superposition 
$A+B$ of different components which scatter elastically off the 
target~\cite{gw} (Fig.~\ref{fig:gwp}b). Since the corresponding elastic 
amplitude is different for each component, the outgoing beam will contain 
a new superposition $\alpha A+\beta B$ of these components and therefore, 
in general, new physical states. These are the dissociation products of the 
projectile. 

Even at very high energy, the above processes are generically soft, i.e., 
the momentum transfer is small and the dissociation products have small 
$p_\perp$. Therefore, no immediate relation to perturbative QCD is apparent. 

\begin{figure}[ht]
\begin{center}
\vspace*{.2cm}
\parbox[b]{4cm}{\psfig{width=4cm,file=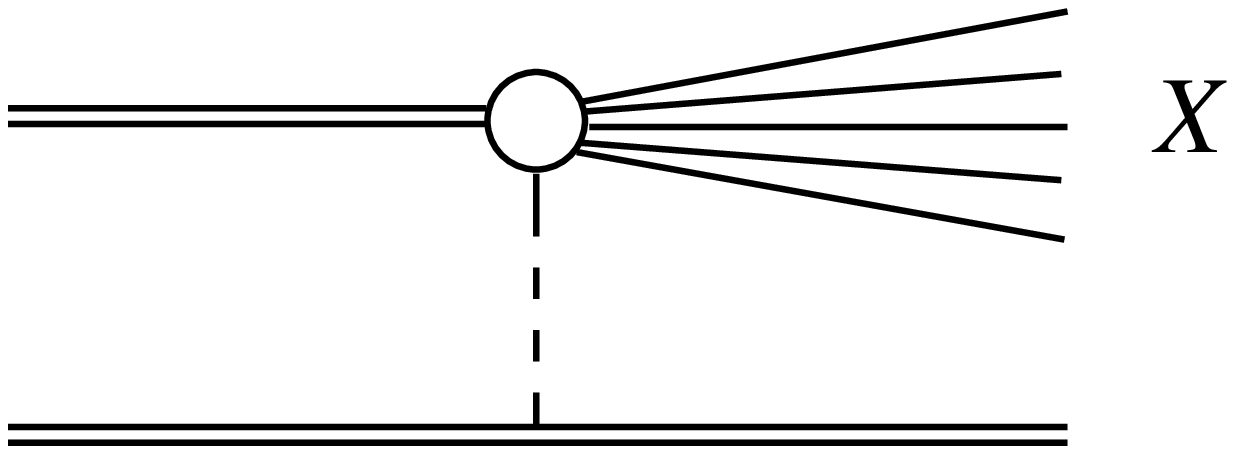}}\\
\end{center}
\refstepcounter{figure}
\label{fig:djp}
{\bf Figure \ref{fig:djp}:} Diffractive dissociation in hadron-hadron 
collisions. If the diffractive final state $X$ contains two high-$p_\perp$ 
jets, the process is called diffractive jet production. 
\end{figure}

In contrast to these soft processes, diffractive DIS is an example of 
hard diffraction, the hard scale being the virtuality $Q^2$ of the exchanged 
photon. However, this is not the first hard diffractive process that was 
observed. Hard diffraction was first investigated in diffractive jet 
production at the CERN S$p\bar{p}$S collider in proton-antiproton 
collisions~\cite{ua8}. Although one of the hadrons escapes essentially 
unscathed, a high-$p_\perp$ jet pair, which is necessarily associated with 
a high virtuality in the intermediate states, is produced in the central 
rapidity range (Fig.~\ref{fig:djp}). The cross section of the process is 
parametrically unsuppressed relative to non-diffractive jet production. 
This seems to contradict a na\"\i ve partonic picture since the colour 
neutrality of the projectile is destroyed if one parton is removed to 
participate in the hard scattering. The interplay of soft and hard physics 
necessary to explain the effect provides one of the main motivations for 
the study of these `hard diffractive' processes. 

\begin{figure}[ht]
\begin{center}
\vspace*{.2cm}
\parbox[b]{10cm}{\psfig{width=10cm,file=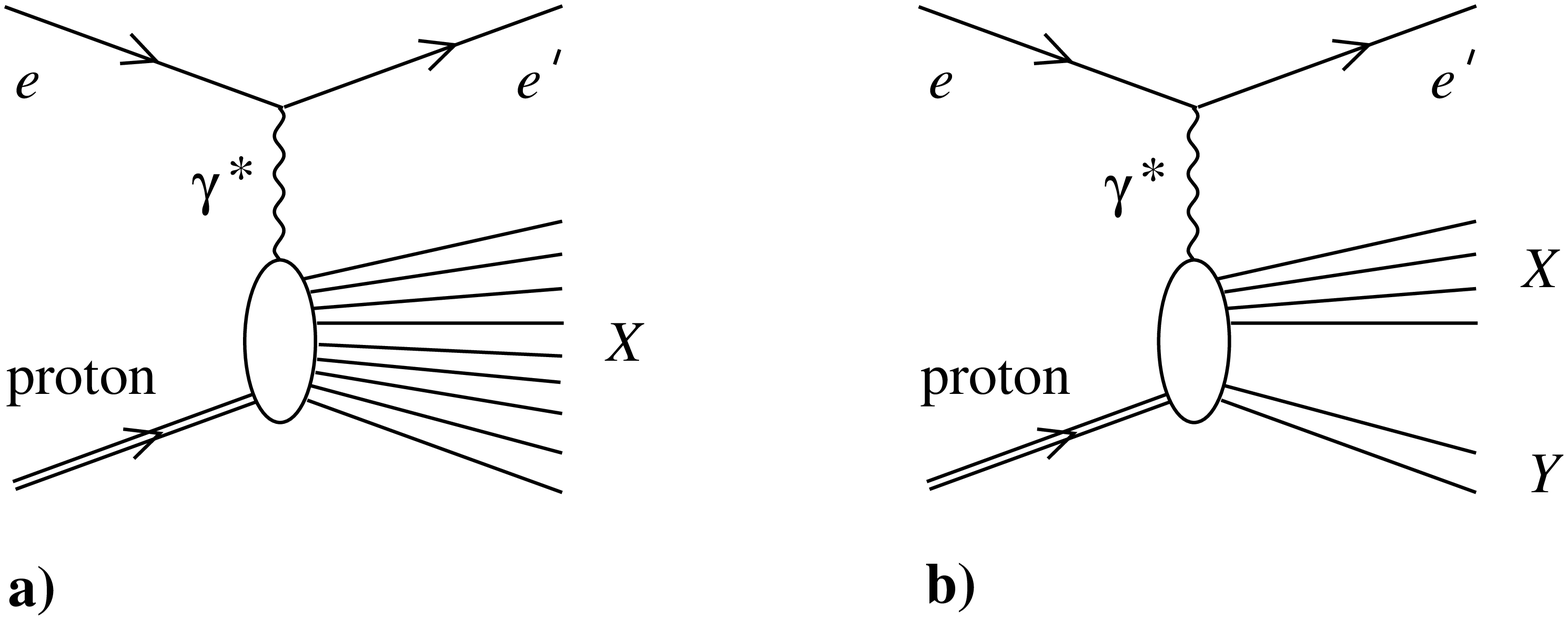}}\\
\end{center}
\refstepcounter{figure}
\label{fig:dis}
{\bf Figure \ref{fig:dis}:} Inclusive (a) and diffractive (b) 
electroproduction. 
\end{figure}

Here, the focus is on diffractive electroproduction, which is another 
example of a hard diffractive process. This process became experimentally 
viable with the advent of the electron-proton collider HERA 
(cf.~\cite{der}), where DIS at 
very small values of the Bjorken variable $x$ can be studied. In the 
small-$x$ or high-energy region, a significant fraction of the observed DIS 
events have a large rapidity gap between the photon and the proton 
fragmentation region~\cite{rg1,rg2}. In contrast to the standard DIS 
process $\gamma^*p\to X$ (Fig.~\ref{fig:dis}a), the relevant reaction 
reads $\gamma^*p\to XY$ (Fig.~\ref{fig:dis}b), where $X$ is a high-mass 
hadronic state and $Y$ is the elastically scattered proton or a low-mass 
excitation of it. Again, these events are incompatible with the na\"\i ve 
picture of a partonic target and corresponding simple ideas about the 
colour flow. Na\"\i vely, the parton struck by the virtual photon destroys 
the colour neutrality of the proton, a colour string forms between struck 
quark and proton remnant, and hadronic activity is expected throughout the 
detector. Nevertheless, the observed diffractive cross section is not power 
suppressed at high virtualities $Q^2$ with respect to standard DIS. 

These notes are organized as follows. In Sect.~\ref{phen}, the basic 
experimental observations and the diffractive structure function $F_2^D$ 
are discussed. Section~\ref{f2th} describes the theory of $F_2^D$, 
emphasizing the semiclassical approach and its relation to the concept of 
diffractive parton distributions. In Sect.~\ref{vm}, some topics in 
diffractive meson production, in particular the factorization of the 
hard amplitude, are described. A brief summary is given in Sect.~\ref{sum}. 

The reader may consult~\cite{hab} for a more detailed review.

\section{Basic observations and the diffractive structure function $F_2^D$}
\label{phen}
In this section, the kinematics of diffractive electroproduction at small 
$x$ is explained in some detail, and the notation conventionally used for 
the description of this phenomenon is introduced. The main experimental 
observations are discussed, and the concept of the diffractive structure 
function, which is widely used in analyses of inclusive diffraction, is 
explained.

\subsection{Kinematics}
To begin, recall the conventional variables for the description of DIS. 
An electron with momentum $k$ collides with a proton with momentum $P$. 
In neutral current processes, a photon with momentum $q$ and virtuality 
$q^2=-Q^2$ is exchanged, and the outgoing electron has momentum $k'=k-q$. 
In inclusive DIS, no questions are asked about the hadronic final state 
$X_W$, which is only known to have an invariant mass square $W^2=(P+q)^2$. 
The Bjorken variable $x=Q^2/(Q^2+W^2)$ characterizes, in the na\"\i ve 
parton model, the momentum fraction of the incoming proton carried by the 
quark that is struck by the virtual photon. If $x\ll 1$, which is the 
relevant region in the present context, $Q$ is much smaller than the photon 
energy in the target rest frame. In this sense, the photon is almost real 
even though $Q^2\gg\Lambda^2$ (where $\Lambda$ is some soft hadronic scale). 
It is then convenient to think in terms of a high-energy $\gamma^*p$ 
collision with centre-of-mass energy $W$. 

\begin{figure}[t]
\begin{center}
\vspace*{.2cm}
\parbox[b]{6cm}{\psfig{width=6cm,file=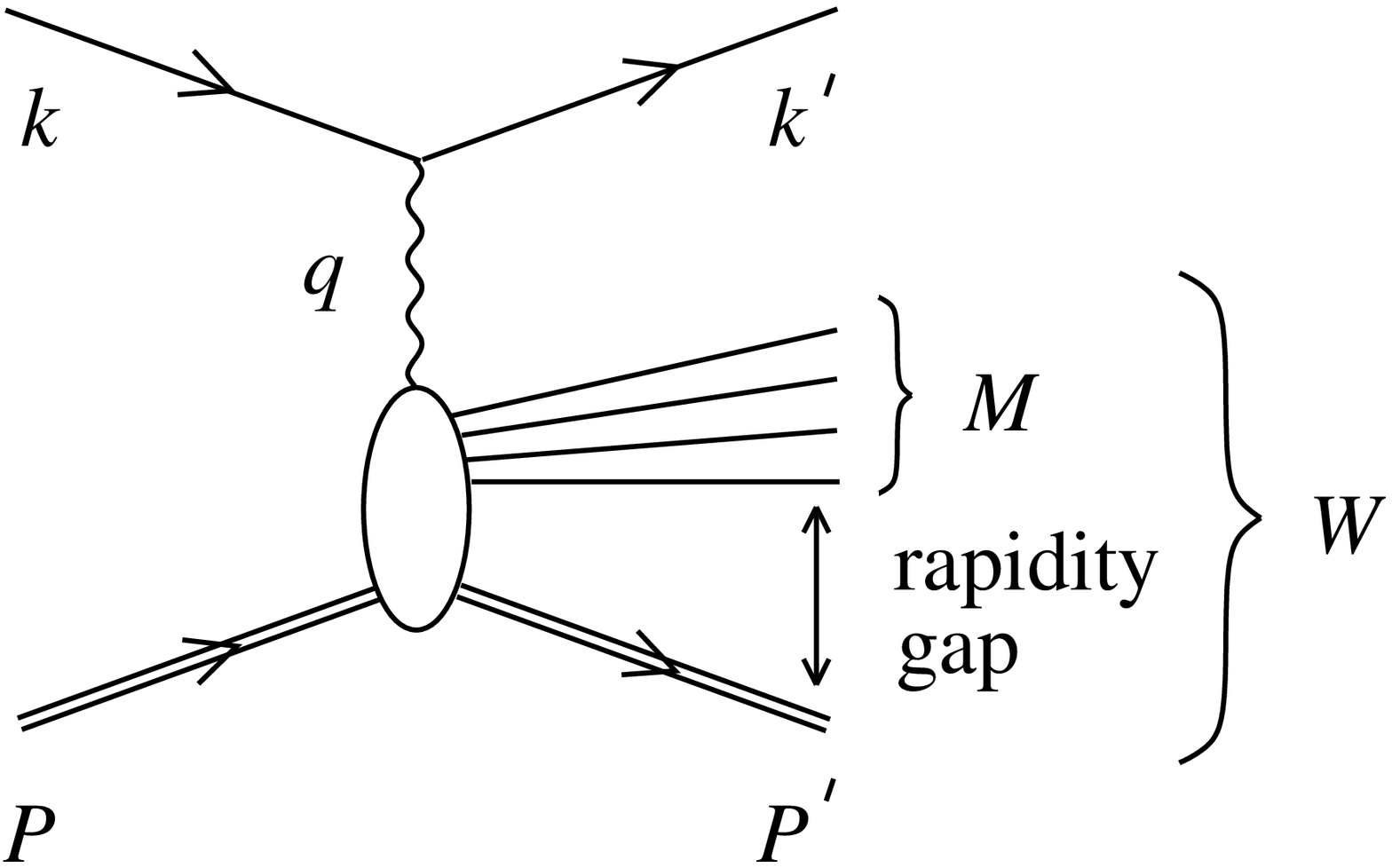}}\\
\end{center}
\refstepcounter{figure}
\label{fig:dep}
{\bf Figure \ref{fig:dep}:} Diffractive electroproduction. The full hadronic 
final state with invariant mass $W$ contains the elastically scattered 
proton and the diffractive state with invariant mass $M$. 
\end{figure}

Loosely speaking, diffraction is the subset of DIS characterized by a 
quasi-elastic interaction between virtual photon and proton. A particularly 
simple definition of diffraction is obtained by demanding that, in the 
$\gamma^*p$ collision, the proton is scattered elastically. Thus, in 
diffractive events, the final state contains the scattered proton with 
momentum $P'$ and a diffractive hadronic state $X_M$ with mass $M$ (see 
Fig.~\ref{fig:dep}). Since diffractive events form a subset of DIS events, 
the total invariant mass of the outgoing proton and the diffractive state 
$X_M$ is given by the standard DIS variable $W$. 

The following parallel description of inclusive and diffractive DIS suggests 
itself. 

In the former, virtual photon and proton collide to form a hadronic 
state $X_W$ with mass $W$. The process can be characterized by the 
virtuality $Q^2$ and the scaling variable $x=Q^2/(Q^2+W^2)$, the momentum 
fraction of the struck quark in the na\"\i ve parton model.

In the latter, a colour neutral cluster of partons is stripped off the 
proton. The virtual photon forms, together with this cluster, a hadronic 
state $X_M$ with mass $M$. The process can be characterized by $Q^2$, as 
above, and by a new scaling variable $\beta=Q^2/(Q^2+M^2)$, the momentum 
fraction of this cluster carried by the struck quark. 

Since diffraction is a subprocess of inclusive DIS, the struck quark from 
the colour neutral cluster also carries a fraction $x$ of the proton 
momentum. Therefore, the ratio $\xi=x/\beta$ characterizes the momentum 
fraction that the proton loses to the colour neutral exchange typical of an 
elastic reaction. This exchanged colour neutral cluster loses a momentum 
fraction $\beta$ to the struck quark that absorbs the virtual photon. As 
expected, the product $x=\beta\xi$ is the fraction of the original proton's 
momentum carried by this struck quark. Since the name pomeron is frequently 
applied to whichever exchange with vacuum quantum numbers dominates the 
high-energy limit, many authors use the notation $x_{\pom}=\xi$, thus 
implying that the proton loses a momentum fraction $\xi$ to the exchanged 
pomeron. 

Therefore, $x$, $Q^2$ and $\beta$ or, alternatively, $x$, $Q^2$ and $\xi$ 
are the main kinematic variables characterizing diffractive DIS. A further 
variable, $t=(P-P')^2$, is necessary if the transverse momenta of the 
outgoing proton and the state $X_M$ relative to the $\gamma^*P$ axis are 
measured. Since the proton is a soft hadronic state, the value of $|t|$ is 
small in most events. The small momentum transferred by the proton also 
implies that $M\ll W$. 

To see this in more detail, introduce light-cone co-ordinates $q_\pm=q_0\pm 
q_3$ and $q_\perp=(q_1,q_2)$. It is convenient to work in a frame where the 
transverse momenta of the incoming particles vanish, $q_\perp=P_\perp=0$. 
Let $\Delta$ be the momentum transferred by the proton, $\Delta=P-P'$, and 
$m_{\pro}^2=P^2=P'^2$ the proton mass squared. For forward scattering, 
$P_\perp'=0$, the relation 
\be
t=\Delta^2=\Delta_+ \Delta_-=-\xi^2m_{\pro}^2
\ee
holds. Since $\xi=(Q^2+M^2)/(Q^2+W^2)$, this means that small $M$ implies 
small $|t|$ and vice versa. Note, however, that the value of $|t|$ is larger 
for non-forward processes, where $t=\Delta_+ \Delta_--\Delta_\perp^2$. 

So far, diffractive events have been characterized as those DIS events 
which contain an elastically scattered proton in their hadronic final 
state. An even more striking feature is the large gap of hadronic activity 
seen in the detector between the scattered proton and the diffractive state 
$X_M$. It will now be demonstrated that this feature, responsible for the 
alternative name `rapidity gap events', is a direct consequence of the 
relevant kinematics. 

Recall the definition of the rapidity $y$ of a particle with momentum 
$k$,
\be
y=\frac{1}{2}\ln\frac{k_+}{k_-}=\frac{1}{2}\ln\frac{k_0+k_3}{k_0-k_3}\,.
\ee
This is a convenient quantity for the description of high-energy collisions 
along the $z$-axis. Massless particles moving along this axis have rapidity 
$-\infty$ or $+\infty$, while all other particles are characterized by some 
finite intermediate value of $y$. 

In the centre-of-mass frame of the $\gamma^*p$ collision, with the $z$-axis 
pointing in the proton beam direction, the rapidity of the incoming 
proton is given by $y_{\pro}=\ln(P_+/m_{\pro})$. At small $\xi$, the 
rapidity of the scattered proton is approximately the same. This is to be 
compared with the highest rapidity $y_{\max}$ of any of the particles in the 
diffractive state $X_M$. Since the total plus component of the 4-momentum
of $X_M$ is given by $(\xi-x)P_+$, and the pion, with mass $m_\pi$, is the 
lightest hadron, none of the particles in $X_M$ can have a rapidity above 
$y_{\max}=\ln((\xi-x)P_+/m_\pi)$. Thus, a rapidity gap of size $\Delta y=\ln 
(m_\pi/(\xi-x)m_{\pro})$ exists between the outgoing proton and the state 
$X_M$. For typical values of $\xi \sim 10^{-3}$ the size of this gap can be 
considerable. 

Note, however, that the term `rapidity gap events' was coined to describe the 
appearance of diffractive events in the HERA frame, i.e., a frame defined by 
the electron-proton collision axis. The rapidity in this frame is, in 
general, different from the photon-proton frame rapidity discussed above. 
Nevertheless, the existence of a gap surrounding the outgoing proton in the 
$\gamma^*p$ frame clearly implies the existence of a similar gap in the 
$ep$ frame. The exact size of the $ep$-frame rapidity gap follows from the 
specific event kinematics. The main conclusion so far is the kinematic 
separation of outgoing proton and diffractive state $X_M$ in diffractive 
events with small $\xi$. 

Without losing any of the qualitative results, the requirement of a final 
state proton $P'$ can be replaced by the requirement of a low-mass hadronic 
state $Y$, well separated from the diffractive state $X_M$. In this case, 
the argument connecting elastically scattered proton and rapidity gap has 
to be reversed: the existence of a gap between $X_M$ and $Y$ becomes the 
distinctive feature of diffraction and, under certain kinematic 
conditions, the interpretation of $Y$ as an excitation of the incoming 
proton, which is now {\it almost} elastically scattered, follows.

\subsection{The main experimental observations}
Rapidity gaps are expected even if in all DIS events a quark is knocked out 
of the proton leaving a coloured remnant. The reason for this is the 
statistical distribution of the produced hadrons, which results in a small 
yet finite probability for final states with little activity in any 
specified detector region. However, the observations described below are 
clearly inconsistent with this explanation of rapidity gap events. 

More than 5\% of DIS events at HERA were found to possess a rapidity 
gap~\cite{rg1,rg2}. The analyses are based on the pseudo-rapidity $\eta=- 
\ln\tan(\theta/2)$, where $\theta$ is the angle of an outgoing particle 
relative to the beam axis. Pseudo-rapidity and rapidity are identical for 
massless particles; the difference between these two quantities is 
immaterial for the qualitative discussion below. 

\begin{figure}[ht]
\begin{center}
\vspace*{.2cm}
\parbox[b]{7cm}{\psfig{width=7cm,file=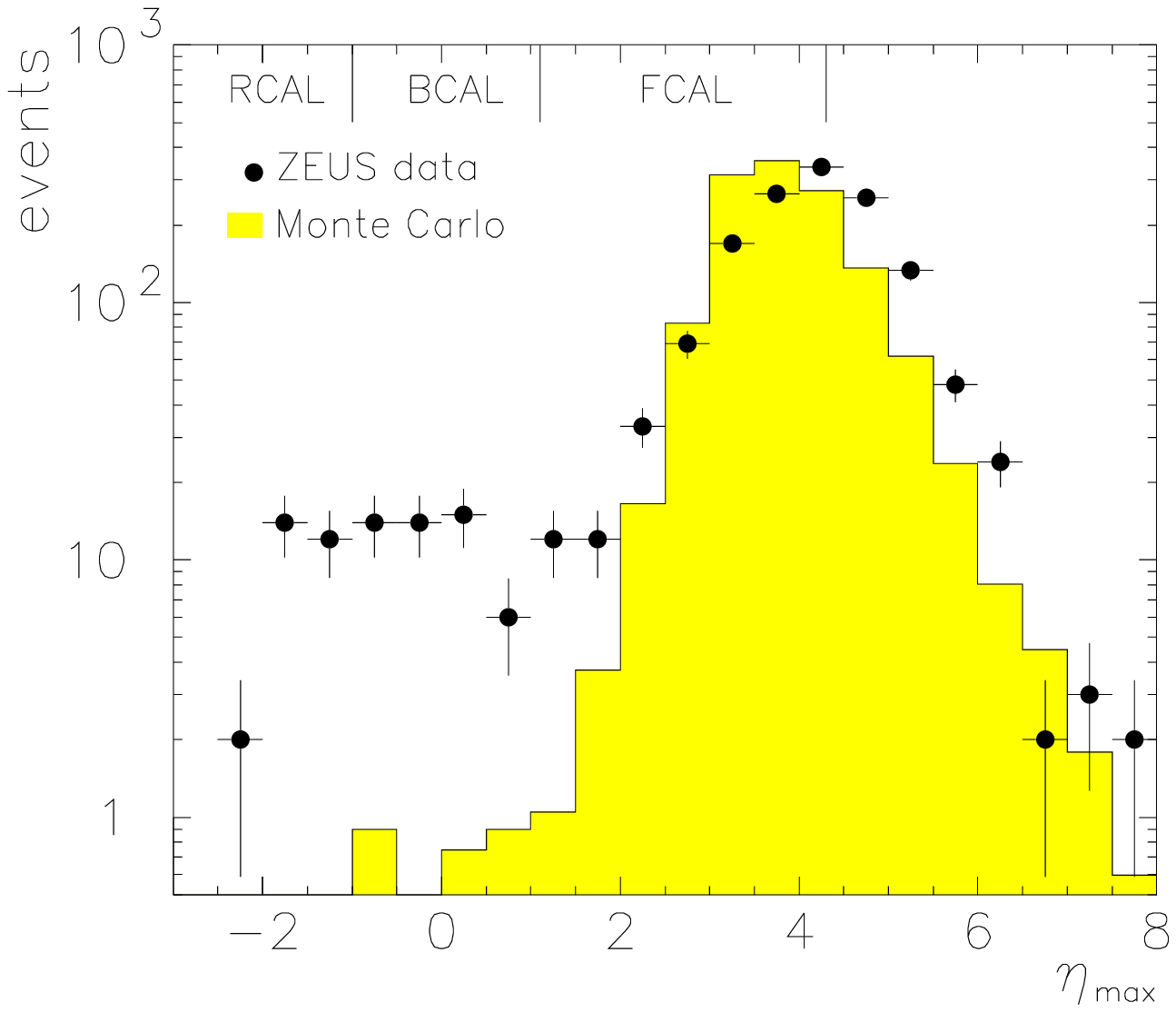}}\\
\end{center}
\refstepcounter{figure}
\label{fig:zeus}
{\bf Figure \ref{fig:zeus}:} Distribution of $\eta_{\max}$, the maximum 
rapidity of a calorimeter cluster in an event, measured at HERA (figure 
from~\cite{rg1}). 
\end{figure}

In the ZEUS analysis, a rapidity $\eta_{\max}$ was defined as the maximum 
rapidity of a calorimeter cluster in an event. A cluster was defined as an 
isolated set of adjacent cells with summed energy higher than 400 MeV. The 
measured $\eta_{\max}$ distribution is shown in Fig.~\ref{fig:zeus}. (Note 
that the smallest detector angle corresponds to $\eta_{\max}=4.3$; larger 
values are an artifact of the clustering algorithm.)

To appreciate the striking qualitative signal of diffraction at HERA, the 
measured $\eta_{\max}$ distribution has to be compared with na\"\i ve 
expectations based on a purely partonic picture of the proton. This is best 
done using a parton-model-based Monte Carlo event generator. The 
corresponding $\eta_{\max}$ distribution, which is also shown in 
Fig.~\ref{fig:zeus}, is strongly suppressed at small $\eta_{\max}$. This 
qualitative behaviour is expected since the Monte Carlo (for more details 
see~\cite{rg1} and refs. therein) starts from a partonic proton, 
calculates the hard process and the perturbative evolution of the QCD 
cascade, and finally models the hadronization using the Lund string model 
(see, e.g.,~\cite{agis}). According to the Lund model, the colour string, 
which connects all final state partons and the coloured proton remnant, 
breaks up via $q\bar{q}$ pair creation, thus producing the observed mesons. 
The rapidities of these particles follow a Poisson distribution, resulting 
in an exponential suppression of large gaps. 

It should be clear from the above discussion that this result is rather 
general and does not depend on the details of the Monte Carlo (note, 
however, the Monte Carlo based approach of~\cite{eir}, 
which allows for colour reconnection). QCD radiation 
tends to fill the rapidity range between the initially struck quark and the 
coloured proton remnant with partons. A colour string connecting these 
partons is formed, and it is highly unlikely that a large gap emerges in 
the final state after the break-up of this string.

However, the data shows a very different behaviour. The expected exponential 
decrease of the event number with $\eta_{\max}$ is observed only above 
$\eta_{\max}\simeq 1.5$; below this value a large plateau is seen. Thus, 
the na\"\i ve partonic description of DIS misses an essential qualitative 
feature of the data, namely, the existence of non-suppressed large 
rapidity gap events. 

To give a more specific discussion of the diffractive event sample, it is 
necessary to define which events are to be called diffractive or rapidity 
gap events. It is clear from Fig.~\ref{fig:zeus} that, on a qualitative 
level, this can be achieved by an $\eta_{\max}$ cut separating the events 
of the plateau. The resulting qualitative features, observed both by the 
ZEUS~\cite{rg1} and H1 collaborations~\cite{rg2}, are the following. 

There exists a large rapidity interval where the $\eta_{\max}$ distribution 
is flat. For high $\gamma^*p$ energies $W$, the ratio of diffractive 
events to all DIS events is approximately independent of $W$. The $Q^2$ 
dependence of this ratio is also weak, suggesting a leading-twist 
contribution of diffraction to DIS. Furthermore, the diffractive mass 
spectrum is consistent with a $1/M^2$ distribution. 

A number of additional remarks are in order. Note first that the 
observation of a flat $\eta_{\max}$ distribution and of a $1/M^2$ spectrum 
are interdependent as long as masses and transverse momenta of final state 
particles are much smaller than $M^2$. To see this, observe that the plus 
component of the most forward particle momentum and the minus component of 
the most backward particle momentum are largely responsible for the total 
invariant mass of the diffractive final state. This gives rise to the 
relation $dM^2/M^2=d\ln M^2 \sim d\eta_{\max}$, which is equivalent to the 
desired result. 

A significant contribution from exclusive vector meson production, e.g., 
the process $\gamma^*p\to\rho\,p$, is present in the rapidity gap event 
sample. A more detailed discussion of corresponding cross sections and of 
relevant theoretical considerations is given in Sect.~\ref{vm}.

\subsection{Diffractive structure function}
The diffractive structure function, introduced in~\cite{ip} and first 
measured by the H1 collaboration~\cite{h1f2d}, is a powerful concept for the 
analysis of data on diffractive DIS. 

Recall the relevant formulae for inclusive DIS. The cross section for the 
process $ep\to eX$ can be calculated if the hadronic tensor, 
\be
W_{\mu\nu}(P,q)=\frac{1}{4\pi}\sum_X<P|j_\nu^\dagger(0)|X><X|j_\mu(0)|P>
(2\pi)^4\delta^4(q+P-p_X)\,,
\ee
is known. Here $j$ is the electromagnetic current, and the sum is over all 
hadronic final states $X$. Because of current conservation, $q\cdot W=W\cdot 
q=0$, the tensor can be decomposed according to 
\be
W_{\mu\nu}(P,q)\!=\!\left(g_{\mu\nu}\!-\!\frac{q_\mu q_\nu}{q^2}\right)
W_1(x,Q^2)\!+\!\left(P_\mu\!+\!\frac{1}{2x}q_\mu\right)\!\!\left(P_\nu\!+\!
\frac{1}{2x}q_\nu\right)W_2(x,Q^2)\,.\label{dec}
\ee
The data is conveniently analysed in terms of the two structure functions 
\bea
F_2(x,Q^2)&=&(P\cdot q)\,W_2(x,Q^2)\\
F_L(x,Q^2)&=&(P\cdot q)\,W_2(x,Q^2)-2xW_1(x,Q^2)\,.
\eea
Introducing the ratio $R=F_L/(F_2-F_L)$, the electron-proton cross section 
can be written as 
\be
\frac{d^2\sigma_{ep\to eX}}{dx\,dQ^2}=\frac{4\pi\aem^2}{xQ^4}\left\{1-y+
\frac{y^2}{2[1+R(x,Q^2)]}\right\}\,F_2(x,Q^2)\,,
\ee
where $y\!=\!Q^2/sx$, and $s$  is the electron-proton centre-of-mass energy 
squared. In the na\"\i ve parton model or at leading order in $\alpha_s$ in 
QCD, the longitudinal structure function $F_L(x,Q^2)$ vanishes, and $R=0$. 
Since $R$ corresponds to the ratio of longitudinal and transverse virtual 
photon cross sections, $\sigma_L/\sigma_T$, it is always positive, and the 
corrections associated with a non-zero $R$ are small at low values of $y$. 

In diffraction, the two additional kinematic variables $\xi$ and $t$ are 
present. However, no additional independent 4-vector is introduced as long 
as the measurement is inclusive with respect to the azimuthal angle of the 
scattered proton. Therefore, the decomposition in Eq.~(\ref{dec}) remains 
valid, and the two diffractive structure functions $F_{2,L}^{D(4)}(x,Q^2, 
\xi,t)$ can be defined. The diffractive cross section reads 
\be
\frac{d^2\sigma_{ep\to epX_M}}{dx\,dQ^2\,d\xi\,dt}\!=\!\frac{4\pi\aem^2}
{xQ^4}\left\{1\!-\!y+\frac{y^2}{2[1+R^{D(4)}(x,Q^2,\xi,t)]}\right\}
F_2^{D(4)}(x,Q^2,\xi,t)\,,\label{d4s}
\ee
where $R^D=F_L^D/(F_2^D-F_L^D)$. In view of the limited precision of the 
data, the dominance of the small-$y$ region, and the theoretical expectation 
of the smallness of $F_L^D$, the corrections associated with a non-zero 
value of $R^D$ are neglected in the following. 

A more inclusive and experimentally more easily accessible quantity can be 
defined by performing the $t$ integration, 
\be
F_2^{D(3)}(x,Q^2,\xi)=\int dt\, F_2^{D(4)}(x,Q^2,\xi,t)\,.
\ee
The main qualitative features of diffractive electroproduction, already 
discussed in the previous section, become particularly apparent if 
the functional form of $F_2^{D(3)}$ is considered (cf. the data in 
Figs.~\ref{h1} and \ref{zeus} in Sect.~\ref{f2th}). The $\beta$ 
and $Q^2$ dependence of $F_2^{D(3)}$ is relatively flat. This corresponds 
to the observations discussed earlier that diffraction is a leading twist 
effect and that the mass distribution is consistent with a $1/M_X^2$ 
spectrum. The energy dependence of diffraction is such that $\xi F_2^{D(3)}
(\xi,\beta,Q^2)$ slowly grows as $\xi\to 0$. This growth and its possible 
relation to the growth of inclusive structure functions at small $x$ and to 
the growth of elastic cross sections at high energy is one of the most
interesting aspects of diffraction (see also~\cite{buch}).

\section{Theoretical approaches to $F_2^D$}\label{f2th}
To understand qualitatively how leading twist diffraction in DIS comes 
about, it is simplest to work in the target rest frame. Diffraction means 
that a hadronic fluctuation of the energetic virtual photon scatters 
off the target proton without exchanging colour. One can then expect the 
photon and the proton to fragment independently, leading to a rapidity gap 
event. 

A particularly simple physical picture is provided by the aligned jet model 
of Bjorken and Kogut~\cite{bk}. In terms of the QCD degrees of freedom,
a closely related formulation of diffraction was given with the two-gluon 
exchange calculations of Nikolaev and Zakharov~\cite{wf}. Since the 
$t$ channel colour singlet exchange is not hard in the bulk of the cross 
section underlying $F_2^D$, the exchange of more than two gluons is not 
suppressed. This can be systematically treated in the semiclassical 
approach~\cite{bh2,bgh}. 

In the Breit frame, where the target proton is fast, diffraction was 
historically described as DIS off a pomeron~\cite{is,dl1}. A more general, 
QCD based concept treating diffraction from a Breit frame point of 
view are the diffractive parton distributions~\cite{vt,bs}. As will be 
discussed in more detail below, the semiclassical approach provides a 
convenient framework in which one can intuitively understand how 
diffractive parton distributions emerge in a target rest calculation.

\subsection{Aligned jet model}
A very simple argument why, even at very high photon virtualities, 
diffractive DIS is largely a soft process was presented by Bjorken and 
Kogut in the framework of their aligned jet model~\cite{bk}.

\begin{figure}[ht]
\begin{center}
\vspace*{.2cm}
\parbox[b]{10cm}{\psfig{width=10cm,file=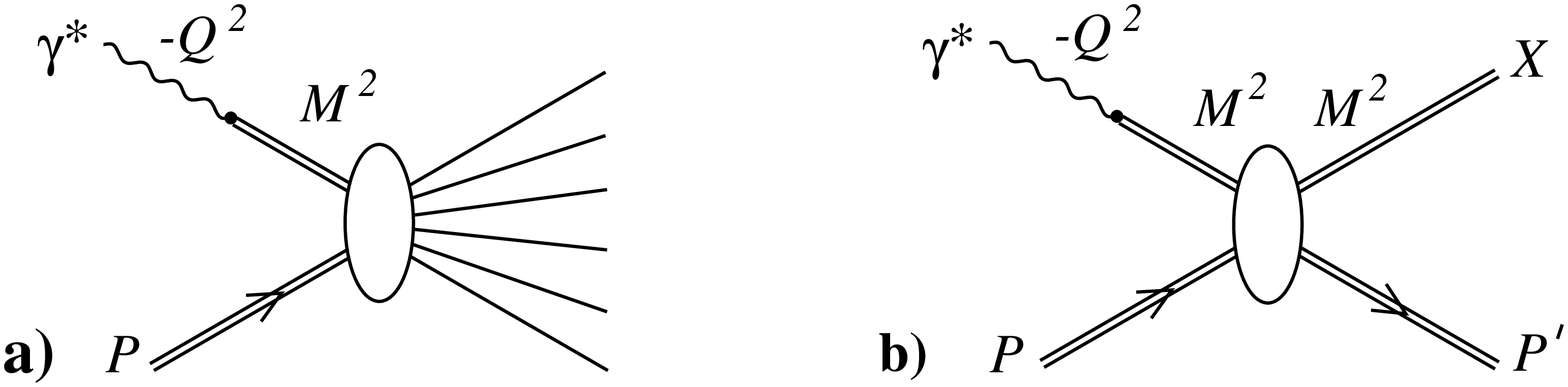}}\\
\end{center}
\refstepcounter{figure}
\label{fig:ajm}
{\bf Figure \ref{fig:ajm}:} Vector meson dominance inspired picture of 
inclusive (a) and diffractive (b) electroproduction.
\end{figure}

The underlying physical picture is based on vector meson dominance ideas. 
The incoming photon fluctuates into a hadronic state with mass $M$, which 
then collides with the target (see Fig.~\ref{fig:ajm}a). The corresponding 
cross section for transverse photon polarization is estimated by 
\be
\frac{d\sigma_T}{dM^2}\sim \frac{dP(M^2)}{dM^2}\cdot\sigma(M^2)\,,
\ee
where the probability for the photon to develop a fluctuation with mass $M$ 
is given by 
\be
dP(M^2)\sim\frac{M^2dM^2}{(M^2+Q^2)^2}\,,
\ee
and $\sigma(M^2)$ is the cross section for this fluctuation to scatter off 
the target. The above expression for $dP(M^2)$ is most easily motivated in 
the framework of old-fashioned perturbation theory, where the energy 
denominator of the amplitude is proportional to the off-shellness of the 
hadronic fluctuation, $Q^2+M^2$. If this is the only source for a $Q^2$ 
dependence, the numerator factor $M^2$ is necessary to obtain a 
dimensionless expression. 

Bjorken and Kogut assume that, for large $M^2$, the intermediate hadronic 
state typically contains two jets and that $\sigma(M^2)$ is suppressed for 
configurations with high $p_\perp$ (the latter effect being now known under 
the name of colour transparency). Consider hadronic fluctuations with a 
certain $M^2$, which, in their respective rest frames, are realized by two 
back-to-back jets. Under the assumption that the probability distribution 
of the direction of the jet axis is isotropic, simple geometry implies that 
aligned configurations, defined by $p_\perp^2<\Lambda^2$ (where $\Lambda^2$ 
is a soft hadronic scale), are suppressed by $\Lambda^2/M^2$. If only such 
configurations are absorbed with a large, hadronic cross section, the 
relations $\sigma(M^2) \sim 1/M^2$ and 
\be
\frac{d\sigma_T}{dM^2}\sim \frac{1}{(M^2+Q^2)^2}\,\label{ajm}
\ee
follow. Thus, the above cross section can be interpreted as the total 
high-energy cross section of target proton and aligned jet fluctuation of 
the photon, i.e., of two soft hadronic objects. Therefore, a similar elastic 
cross section is expected, $\sigma^D_T\sim\sigma_T$ (cf. 
Fig.~\ref{fig:ajm}b). The resulting diffractive structure function reads 
\be
F_2^{D(3)}(\xi,\beta,Q^2)\sim \beta/\xi\,.\label{ajmf2d}
\ee

It is interesting that the very simple arguments outlined above capture 
two important features of the HERA data: the leading-twist nature of 
diffraction and the approximate $1/\xi$ behaviour of $F_2^D$.

\subsection{Two-gluon exchange}
\begin{figure}[ht]
\begin{center}
\vspace*{.2cm}
\parbox[b]{6.3cm}{\psfig{width=6.3cm,file=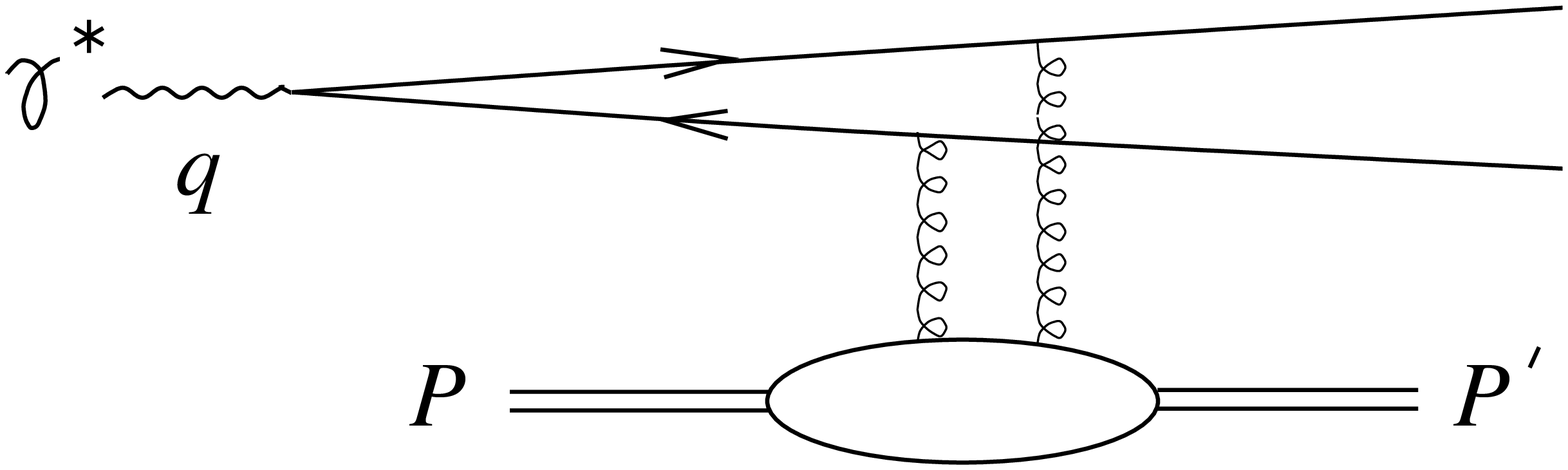}}
\end{center}
\refstepcounter{figure}
\label{fig:gg}
{\bf Figure \ref{fig:gg}:} Diffractive electroproduction in the two-gluon 
exchange model. 
\end{figure}

The simplest way to formulate the above intuitive picture, where a 
had\-ro\-nic fluctuation of the photon scatters off the proton, in 
perturbative QCD is via two-gluon exchange. At leading order, the photon 
fluctuates in a $q\bar{q}$ pair and two gluons forming a colour singlet 
are exchanged in the $t$ channel (see Fig.~\ref{fig:gg}). A corresponding 
calculation was first performed by Nikolaev and Zakharov~\cite{wf}, where 
the probability for the photon to fluctuate in a $q\bar{q}$ pair was 
described by the square of the light-cone wave function of the photon. The 
scattering of the $q\bar{q}$ pair off the proton was parametrized using 
the dipole cross section $\sigma(\rho)$, where $\rho$ is the transverse 
distance between quark and antiquark when they hit the target. Here, the 
corresponding calculations will not be described (see, however, the recent 
review of diffraction~\cite{mw} emphasizing two-gluon exchange). Note that 
two-gluon exchange results can be recovered from the semiclassical 
calculation described below if a Taylor expansion in the external colour 
field is performed. 

Gluon radiation, i.e., the fluctuation of the incoming photon in a 
$q\bar{q}g$ state was considered in~\cite{nz}. 

For further interesting work related to the two-gluon exchange 
approximation for $F_2^D$ see, e.g.,~\cite{bw,bhma,bekw} (cf.~\cite{sal} 
for new developments in the BFKL resummation method). Note also the 
discussion of diffractive processes in the colour-dipole 
approach~\cite{bia} and the combined analysis of diffractive and 
inclusive DIS in this framework~\cite{pes}. 

The main shortcoming of the two-gluon approach is the lacking justification 
of perturbation theory. As should be clear from the aligned jet model and 
as will be discussed in a more technical way below, the diffractive 
kinematics is such that the $t$ channel colour singlet exchange does not 
feel the hard scale of the initial photon. Thus, more than two gluons can 
be exchanged without suppression by powers of $\alpha_s$. 

Note that this is different in certain more exclusive processes, such as
vector meson production, where it has been shown that the dynamics of the 
$t$ channel exchange is governed by a hard, perturbative scale (cf. 
Sect.~\ref{vm}).

\subsection{Semiclassical approach}\label{sa}
In this approach, the interaction with the target is modelled as the 
scattering off a superposition of soft target colour fields, which, in the 
high-energy limit, can be calculated in the eikonal 
approximation~\cite{nac}. Diffraction occurs if both the target and the 
partonic fluctuation of the photon remain in a colour singlet state. 

The amplitude for an energetic parton to scatter off a given colour field 
configuration is a fundamental building block in the semiclassical 
approach. The essential assumptions are the softness and localization of 
the colour field and the very large energy of the scattered parton. 

\begin{figure}[ht]
\begin{center}
\vspace*{.2cm}
\parbox[b]{4.5cm}{\psfig{width=4.5cm,file=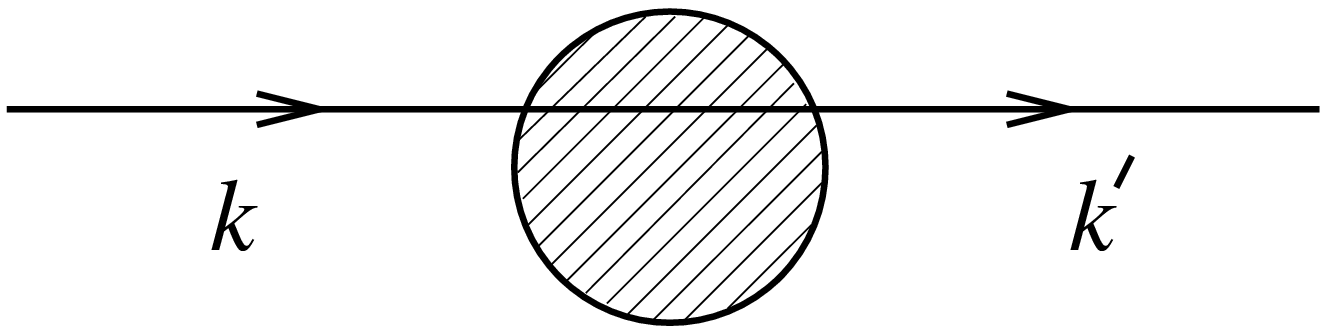}}\\
\end{center}
\refstepcounter{figure}
\label{fig:qs}
{\bf Figure \ref{fig:qs}:} Scattering of a quark off the target colour 
field. 
\end{figure}

The relevant physical situation is depicted in Fig.~\ref{fig:qs}, where the 
blob symbolizes the target colour field configuration. Consider first the 
case of a scalar quark that is minimally coupled to the gauge field via the 
Lagrangian 
\be
{\cal L}_{\sca}=\left(D_\mu \Phi\right)^*\left(D^\mu\Phi\right)\qquad
\mbox{with}\qquad D_\mu=\partial_\mu+igA_\mu\,.
\ee
In the high-energy limit, where the plus component of the quark momentum 
becomes large, the amplitude of Fig.~\ref{fig:qs} then reads 
\be
i2\pi\delta(k_0'-k_0)T=2\pi\delta(k_0'-k_0)2k_0\left[\tilde{U}
(k_\perp'-k_\perp)-(2\pi)^2\delta^2(k_\perp'-k_\perp)\right]\,.\label{qs} 
\ee
It is normalized as is conventional for scattering processes off a fixed 
target. The expression in square brackets is the Fourier transform of the 
impact parameter space amplitude, $U(x_\perp)-1$, where 
\be
U(x_\perp)=P\exp\left(-\frac{ig}{2}\int_{-\infty}^{\infty}A_-(x_+,x_\perp)
dx_+\right)\label{um}
\ee
is the non-Abelian eikonal factor. The unit matrix $1\!\in SU(N_c)$, with 
$N_c$ the number of colours, subtracts the field independent part, and the 
path ordering operator $P$ sets the field at smallest $x_+$ to the 
rightmost position. 

This formula was derived by many authors~\cite{bks,css,nac}. A derivation 
based on the summation of diagrams of the type shown in Fig.~\ref{fig:gsu}, 
is given in the Appendix of~\cite{hab}. 

The amplitude of Eq.~(\ref{qs}) is easily generalized to the case of a 
spinor quark or an energetic gluon. In the the high-energy limit, 
helicity-flip or polarization-flip contributions are suppressed. 

\begin{figure}[ht]
\begin{center}
\vspace*{.2cm}
\parbox[b]{5.4cm}{\psfig{width=5.4cm,file=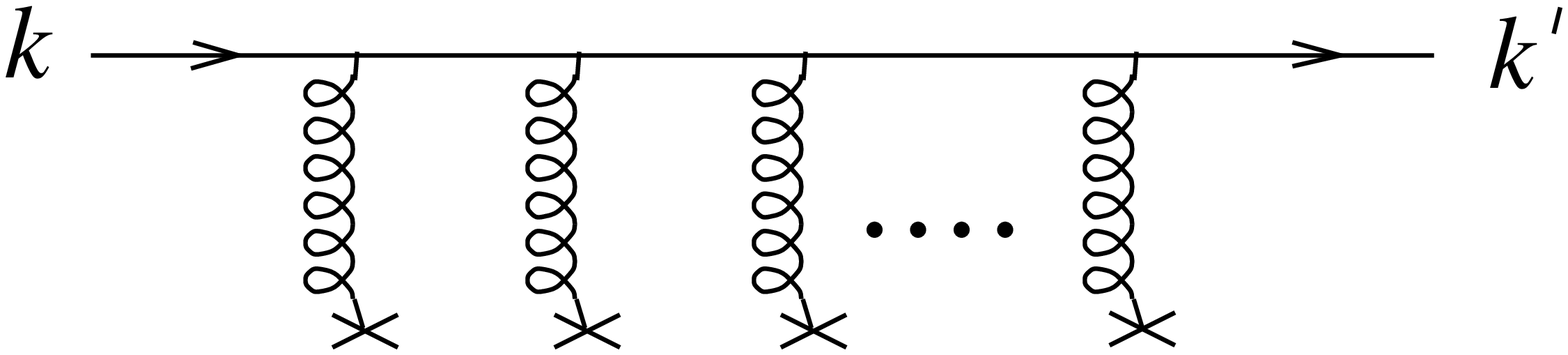}}\\
\end{center}
\refstepcounter{figure}
\label{fig:gsu}
{\bf Figure \ref{fig:gsu}:} Typical diagrammatic contribution to the eikonal 
amplitude, Eq.~(\ref{qs}). Attachments of gluon lines with crosses 
correspond to vertices at which the classical external field appears. 
\end{figure}

The eikonal approximation can be used for the calculation of the amplitude 
for $q\bar{q}$ pair production off a given target colour field~\cite{bh2}. 
Both diffractive and inclusive cross sections are obtained from the same 
calculation, diffraction being defined by the requirement of colour 
neutrality of the produced pair. 

\begin{figure}[ht]
\begin{center}
\vspace*{.2cm}
\parbox[b]{5.7cm}{\psfig{width=5.7cm,file=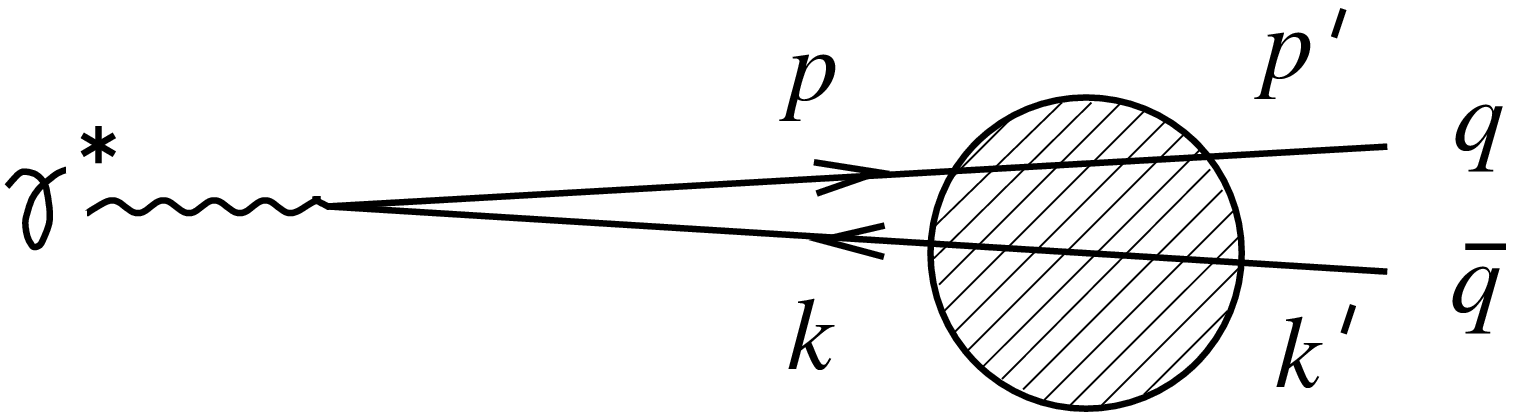}}\\
\end{center}
\refstepcounter{figure}
\label{fig:qq}
{\bf Figure \ref{fig:qq}:} Electroproduction of a $q\bar{q}$ pair off the 
target colour field.
\end{figure}

The process is illustrated in Fig.~\ref{fig:qq}. The corresponding $T$ 
matrix element has three contributions, $T=T_{q\bar{q}}+T_q+T_{\bar{q}}$, 
where $T_{q\bar{q}}$ corresponds to both the quark and antiquark interacting 
with the field, while $T_q$ and $T_{\bar{q}}$ correspond to only one of the 
partons interacting with the field. 

Let $V_q(p',p)$ and $V_{\bar{q}}(k',k)$ be the effective vertices for an 
energetic quark and antiquark interacting with a soft gluonic field. For 
quarks with charge $e$, one has
\bea
i2\pi\delta(k_0'+k_0-q_0)T_{q\bar{q}}&&\label{tqq}\\
&&\hspace*{-3cm}=\,ie\int\frac{d^4k}{(2\pi)^4}\bar{u}_{s'}(p')V_q(p',p)
\frac{i}{\ppl-m}\epsilons(q)\frac{i}{-\ks-m}V_{\bar{q}}(k,k')v_{r'}(k')
\,,\nonumber
\eea
where $q=p+k$ by momentum conservation, $\epsilon(q)$ is the polarization 
vector of the incoming photon, and $r',s'$ label the spins of the outgoing 
quarks. 

The propagators in Eq.~(\ref{tqq}) can be treated in a high-energy 
approximation. In a co-ordinate system where the photon momentum 
is directed along the $z$-axis, the large components are $p_+$ and $k_+$. 
It is convenient to introduce, for each vector $k$, a vector $\bar{k}$ 
whose minus component satisfies the mass shell condition, $\bar{k}_-=
(k_{\perp}^2+m^2)/k_+$, while the other components are identical to 
those of $k$. The propagators in Eq.~(\ref{tqq}) can be rewritten 
according to 
\be
{1\over \ppl - m}=\frac{\sum_s u_s(\bar{p}) \bar{u}_s(\bar{p})}
{p^2-m^2} + \frac{\gamma_+}{2 p_+}\label{prop2}
\ee
and an analogous identity for the propagator with momentum $k$. In the 
high-energy limit, the term proportional to $\gamma_+$ in 
Eqs.~(\ref{prop2}) can be dropped. 

After inserting Eqs.~(\ref{prop2}) into Eq.~(\ref{tqq}), the relation 
\be
\bar{u}_{s'}(p')V_q(p',p)u_s(p)=2\pi\delta(p_0'-
p_0)2p_0\left[\tilde{U}(p_\perp'\!-\!p_\perp)-(2\pi)^2\delta^2(p_\perp'\!-
\!p_\perp)\right]\delta_{ss'}\,,\label{v1}
\ee
which corresponds to Eq.~(\ref{qs}), is applied. The vertex $V_{\bar{q}}
(k,k')$ is treated analogously. Writing the loop integration as $d^4k= 
(1/2)dk_+dk_-d^2k_\perp$ and using the approximation $\delta(l_0)\simeq 
2\delta(l_+)$ for the energy $\delta$-functions, the $k_+$ integration 
becomes trivial. The $k_-$ integral is done by closing the integration 
contour in the upper or lower half of the complex $k_-$ plane. The result 
reads 
\bea
T_{q\bar{q}} &=&  -{ie\over 4\pi^2}\ q_+
  \int d^2k_{\perp}\ {\alpha(1-\alpha)\over N^2 + k_{\perp}^2}
  \ \bar{u}_{s'}(\bar{p})\epsilons(q)v_{r'}(\bar{k})\label{tqbq}\\
&&\hspace*{-1.5cm}\times \left[\tilde{U}(p_\perp'-p_\perp)-(2\pi)^2
\delta^2(p_\perp'-p_\perp)\right]\,\,\left[\tilde{U}^\dagger(k_\perp-
k_\perp')-(2\pi)^2\delta^2(k_\perp'-k_\perp)\right]\nonumber
\eea
where $p'_+=(1-\alpha)q_+$, $k'_+=\alpha q_+$, $N^2=\alpha(1-\alpha)Q^2+ 
m^2$. Thus, $\alpha$ and $1\!-\!\alpha$ characterize the fractions of the 
photon momentum carried by the two quarks, while $(N^2+k_\perp^2)$ measures 
the off-shellness of the partonic fluctuation before it hits the target. In 
the following we set $m=0$. 

Adding $T_q$ and $T_{\bar{q}}$ and introducing the fundamental function 
\be
W_{x_\perp}(y_\perp)=U(x_\perp)U^\dagger(x_\perp+y_\perp)-1\,,\label{wdef} 
\ee
which encodes all the information about the external field, the complete 
amplitude can eventually be given in the form 
\be
T = -{ie\over 4\pi^2}\ q_+
  \int d^2k_{\perp}\ {\alpha(1-\alpha)\over N^2 + k_{\perp}^2}
  \ \bar{u}_{s'}(\bar{p})\epsilons(q)v_{r'}(\bar{k})
\int_{x_\perp}e^{-i\Delta_\perp x_\perp}
\tilde{W}_{x_\perp}(k_\perp'-k_\perp)\,,\label{tfi}
\ee
where $\tilde{W}_{x_\perp}$ is the Fourier transform of $W_{x_\perp}
(y_\perp)$ with respect to $y_\perp$, and $\Delta_\perp=k_\perp'+p_\perp'$ 
is the total transverse momentum of the final $q\bar{q}$ state. 

{}From the above amplitude, the transverse and longitudinal virtual photon 
cross sections are calculated in a straightforward manner using the 
explicit formulae for $\bar{u}_{s'}(\bar{p})\epsilons(q)v_{r'}(\bar{k})$. 
Summing over all $q\bar{q}$ colour 
combinations, as appropriate for the inclusive DIS cross section, the 
following result is obtained, 
\bea
\frac{d\sigma_L}{d\alpha\,dk_\perp'^2}&=&
{2e^2 Q^2\over (2\pi)^6}(\alpha(1-\alpha))^2 \int_{x_{\perp}} 
\left|\int d^2 k_{\perp} \frac{\tilde{W}_{x_{\perp}}(k_{\perp}'-k_{\perp})}
{N^2 + k_{\perp}^2}\right|^2\,,\label{dsl}
\\ \nonumber\\
\frac{d\sigma_T}{d\alpha\,dk_\perp'^2}&=&{e^2\over 2(2\pi)^6 }(\alpha^2 + 
(1-\alpha)^2)\int_{x_{\perp}} \left|\int d^2 k_\perp \frac{k_\perp 
\tilde{W}_{x_\perp}(k_\perp'-k_\perp)}{N^2 + k_\perp^2}\right|^2\!\!.
\label{dst}
\eea
The contraction of the colour indices of the two $W$ matrices is implicit. 

An explicit calculation shows (see, e.g.,~\cite{hab}) that the leading 
contributions at high $Q^2$ are 
\be
\sigma_L={e^2\over 6\pi^2 Q^2}\ \int_{x_{\perp}} 
          \left|\partial_{\perp}W_{x_{\perp}}(0)\right|^2\label{slt}
\ee
and
\bea
\sigma_T&=&{e^2\over 6\pi^2 Q^2}\left(\ln\frac{Q^2}{\mu^2}-1\right)\ 
\int_{x_{\perp}}\left|\partial_{\perp}W_{x_{\perp}}(0)\right|^2\label{stt}
\\ \nonumber \\
&&+{e^2\over (2\pi)^6 }\int_0^{\mu^2/Q^2}d\alpha\int
d k_\perp'^2 \int_{x_{\perp}} \left|\int d^2 k_\perp \frac{k_\perp 
\tilde{W}_{x_\perp}(k_\perp'-k_\perp)}{N^2 + k_\perp^2}\right|^2\!\!,
\nonumber
\eea
where $\Lambda^2\ll\mu^2\ll Q^2$. 

The resulting physical picture can be summarized as follows. For 
longitudinal photon polarization, the produced $q\bar{q}$ pair has small 
transverse size and shares the photon momentum approximately equally. Only 
the small distance structure of the target colour field, characterized by 
the quantity $|\partial_{\perp}W_{x_{\perp}}(0)|^2$, is tested. For 
transverse photon polarization, an additional leading twist contribution 
comes from the region where $\alpha$ or $1\!-\!\alpha$ is small and 
$k_\perp'^2\sim\Lambda^2$. In this region, the $q\bar{q}$ pair penetrating 
the target has large transverse size, and the large distance 
structure of the target colour field, characterized by the function 
$W_{x_\perp}(y_\perp)$ at large $y_\perp$, is tested. This physical 
picture, known as the aligned jet model, was introduced in~\cite{bk} on a 
qualitative level and was used more recently for a quantitative discussion 
of small-$x$ DIS in~\cite{wf}. 

Diffractive cross sections are now derived by introducing a colour singlet 
projector into the underlying amplitude, i.e., by the substitution 
\be
\mbox{tr}\left(W_{\x}(\y)W_{\x}^{\dagger}(\y')\right)\rightarrow\frac{1}
{N_c}\mbox{tr}W_{\x}(\y)\mbox{tr}W_{\x}^{\dagger}(\y')\label{wsubs}
\ee
in Eqs.~(\ref{dsl}) and (\ref{dst}). This change of the colour structure 
has crucial consequences. 

Firstly, the longitudinal cross section vanishes at leading twist (cf. 
Eq.~(\ref{slt})) since the derivative $\partial_{\perp}W_{x_{\perp}}(0)$ 
is in the Lie-algebra of $SU(N_c)$, and therefore tr$\,\partial_{\perp} 
W_{x_{\perp}}(0)=0$. 

Secondly, for the same reason the ln$Q^2$ term in the transverse cross 
section, given by Eq.~(\ref{stt}), disappears. The whole cross section is 
dominated by the endpoints of the $\alpha$ integration, i.e., the aligned 
jet region. At leading order in $1/Q^2$, the diffractive cross 
sections read
\bea
\sigma^D_L&\!=\!&0\\ \nopagebreak
\sigma^D_T&\!=\!&{e^2\over (2\pi)^6 N_c}\int_0^\infty d\alpha\int
d k_\perp'^2 \int_{x_{\perp}} \left|\int d^2 k_\perp \frac{k_\perp 
\mbox{tr}\tilde{W}_{x_\perp}(k_\perp'-k_\perp)}{N^2 + k_\perp^2}
\right|^2\!\!.\label{slst}
\eea
The cutoff of the $\alpha$ integration, $\mu^2/Q^2$, has been dropped 
since, due to the colour singlet projection, the integration is 
automatically dominated by the soft endpoint. 

In summary, the leading-twist cross section for small-$x$ DIS receives 
contributions from both small- and large-size $q\bar{q}$ pairs, the latter 
corresponding to aligned jet configurations. The requirement of colour 
neutrality in the final state suppresses the small-size contributions. 
Thus, leading twist diffraction is dominated by the production of pairs 
with large transverse size testing the non-perturbative large-distance 
structure of the target colour field.

\subsection{Diffractive parton distributions}
The basic theoretical ideas are due to Trentadue and Veneziano, who 
proposed to parametrize semi-inclusive hard processes in terms of 
`fracture functions'~\cite{vt}, and to Berera and Soper, who defined 
similar quantities for the case of hard diffraction~\cite{bs} and coined 
the term `diffractive parton distributions'. The following discussion is 
limited to the latter, more specialized framework. 

Recall first that a conventional parton distribution $f_i(y)$ describes 
the probability of finding, in a fast moving proton, a parton $i$ with 
momentum fraction $y$. 

In short, diffractive parton distributions are conditional probabilities. 
A diffractive parton distribution $df^D_i(y,\xi,t)/d\xi\,dt$ describes the 
probability of finding, in a fast moving proton, a parton $i$ with momentum 
fraction $y$, under the additional requirement that the proton 
remains intact while being scattered with invariant momentum transfer $t$ 
and losing a small fraction $\xi$ of its longitudinal momentum. Thus, the 
corresponding $\gamma^*p$ cross section can be written as~\cite{bs1} 
\be
\frac{d\sigma(x,Q^2,\xi,t)^{\gamma^*p\to p'X}}{d\xi\,dt}=\sum_i\int_x^\xi 
dy\,\hat{\sigma}(x,Q^2,y)^{\gamma^*i}\left(\frac{df^D_i(y,\xi,t)}{d\xi\,dt}
\right)\, ,\label{sx}
\ee
where $\hat{\sigma}(x,Q^2,y)^{\gamma^*i}$ is the total cross section for 
the scattering of a virtual photon characterized by $x$ and $Q^2$ and 
a parton of type $i$ carrying a fraction $y$ of the proton momentum. The 
above factorization formula holds in the limit $Q^2\to\infty$ with $x$, 
$\xi$ and $t$ fixed. 

At leading order and in the case of transverse photon polarization, only 
the quark distribution contributes. For one quark flavour with one unit of 
electric charge, the well-known partonic cross section reads 
\be
\hat{\sigma}_T(x,Q^2,y)^{\gamma^*q}=\frac{\pi e^2}{Q^2}\,\delta(1-y/x)\,,
\ee
giving rise to the diffractive cross section 
\be
\frac{d\sigma(x,Q^2,\xi,t)^{\gamma^*p\to p'X}}{d\xi\,dt}=\frac{2 \pi e^2}
{Q^2}\,\frac{x\,df^D_q(x,\xi,t)}{d\xi\,dt}\, ,
\ee
where the factor $2$ is introduced to account for the antiquark 
contribution. 

As in inclusive DIS, there are infrared divergences in the partonic cross 
sections and ultraviolet divergences in the parton distributions. 
They are conveniently renormalized with the $\overline{\mbox{MS}}$ 
prescription, which introduces the scale $\mu$ as a further argument. The 
distribution functions then read $f_i(x,\mu^2)$ and $df^D_i(x,\xi,t,\mu^2) 
/d\xi dt$ in the inclusive and diffractive case respectively. 

Accordingly, Eq.~(\ref{sx}) has to be read in the $\overline{\mbox{MS}}$ 
scheme, with a $\mu$ dependence appearing both in the parton distributions 
and in the partonic cross sections. The claim that Eq.~(\ref{sx}) holds to 
all orders implies that these $\mu$ dependences cancel, as is well known 
in the case of conventional parton distributions. Since the partonic cross 
sections are the same in both cases, the diffractive distributions obey 
the usual Altarelli-Parisi evolution equations, 
\be
\frac{d}{d(\ln\mu^2)}\,\frac{df^D_i(x,\xi,t,\mu^2)}{d\xi\,dt}=\sum_j 
\int_x^\xi\frac{dy}{y}P_{ij}(x/y)\frac{df^D_j(y,\xi,t,\mu^2)}{d\xi\,dt}\,. 
\ee
with the ordinary splitting functions $P_{ij}(x/y)$.

Thus, for the analysis of diffractive DIS, it is essential to gain 
confidence in the validity of the factorization formula Eq.~(\ref{sx}). 
Berera and Soper first pointed out~\cite{bs1} that such a factorization 
proof could be designed along the lines of related results for other QCD 
processes (see~\cite{cssr} for a review). Proofs were given by Grazzini, 
Trentadue and Veneziano in the framework of a simple scalar 
model~\cite{gtv} and by Collins in full QCD~\cite{cfa}. 

The concept of diffractive parton distributions is more rigorous but less 
predictive than the older, widely used method of parton distributions of 
the pomeron~\cite{is,dl1,pp}. This method originates in the observation 
that diffractive 
DIS can be understood as the soft high-energy scattering of the hadronic 
fluctuation of the photon and the target proton. Since a large sample of 
hadronic cross sections can be consistently parametrized using the concept 
of the Donnachie-Landshoff or soft pomeron~\cite{dl}, it is only natural to 
assume that this concept can also be used in the present case. 
A more direct way of applying the concept of the soft pomeron to the 
phenomenon of hard diffraction was suggested by Ingelman and Schlein in 
the context of diffractive jet production in hadronic collisions~\cite{is}. 
Their idea of a partonic structure of the pomeron, which can be tested in 
hard processes, applies to the case of diffractive DIS as well~\cite{dl1}. 
Essentially, one assumes that the pomeron can, like a real hadron, be 
characterized by a parton distribution. This distribution factorizes from 
the pomeron trajectory and the pomeron-proton vertex, which are both 
obtained from the analysis of purely soft hadronic reactions. The above 
non-trivial assumptions, which have not been derived from QCD, are often 
referred to as `Regge hypothesis' or `Regge factorization'.

\subsection{Diffractive parton distributions in the semiclassical 
approach}
It will be shown how the semiclassical calculation can be factorized into 
hard and soft part and how a model for diffractive parton densities 
naturally arises from the soft part of the semiclassical 
calculation~\cite{h}. 

\begin{figure}[ht]
\begin{center}
\vspace*{.2cm}
\parbox[b]{7cm}{\psfig{width=7cm,file=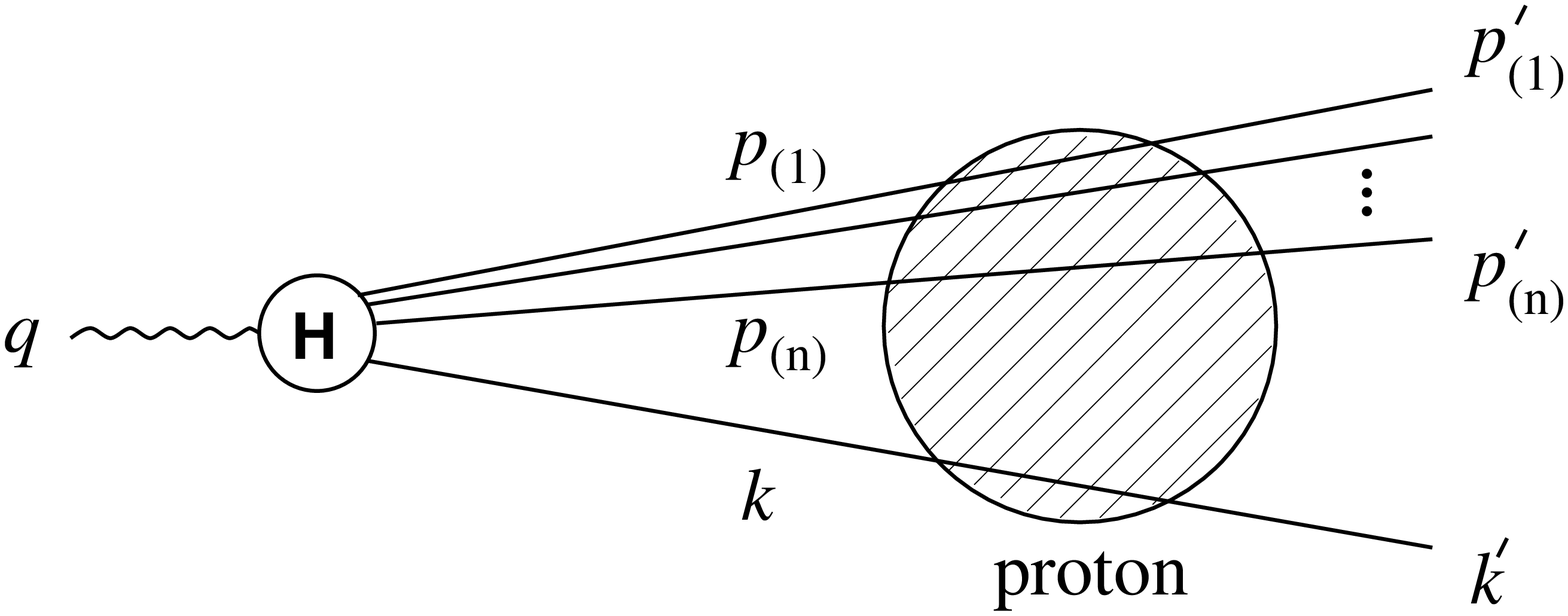}}\\
\end{center}
\refstepcounter{figure}
\label{fig:fac}
{\bf Figure \ref{fig:fac}:} Hard diffractive process in the proton rest 
frame. The soft parton with momentum $k$ is responsible for the leading 
twist behaviour of the cross section. 
\end{figure}

For simplicity, consider the fluctuation of the photon into a set of 
scalar partons which interact independently with the proton colour field 
(see Fig.~\ref{fig:fac}).

Assume furthermore that the transverse momenta $p_{(j)\perp}' (j=1...n)$ are 
hard, i.e. ${\cal O}(Q)$. A leading twist contribution to diffraction can 
arise only if the transverse momentum of the remaining parton is small,
$k_\perp'\sim \Lambda_{\mbox{\scriptsize QCD}}$. 

The standard cross section formula for the scattering off a static external 
field reads
\be
d\sigma=\frac{1}{2q_0}\,|T|^2\,2\pi\delta(q_0-q_0')\,dX^{(n+1)}
\,,\quad\mbox{where}\quad q'=k'+\mbox{$\sum$}p_{(j)}'\, .
\ee
All momenta are given in the proton rest frame, $T$ is the amplitude 
corresponding to Fig.~\ref{fig:fac}, and $dX^{(n+1)}$ is the usual phase 
space element for $n+1$ particles.

The colour rotation experienced by a each parton penetrating the external 
field is described by an eikonal factor. The resulting amplitude is given by
\[
\hspace*{-1.5cm}i\,2\pi\delta(q_0-q_0')\,T\!=\!\int\!T_H\,\left(\frac{i}
{k^2}\,2\pi\delta(k_0'-k_0)\,2k_0\,\tilde{U}(k_\perp'-k_\perp)\right)
\]
\be
\hspace*{1cm}\times\prod_j\!\left(\frac{i}{p_{(j)}^2}
\,2\pi\delta(p_{(j)0}'-p_{(j)0})\,2p_{(j)0}\,\tilde{U}(p_{(j)\perp}'-
p_{(j)\perp})\,\frac{d^4p_{(j)}}{(2\pi)^4}\right)\,,
\ee
where $T_H$ stands for the hard part of the diagram in Fig.~\ref{fig:fac}.

The integrations over the light-cone components $p_{(j)+}$ can be performed 
using the energy $\delta$-functions. The $p_{(j)-}$-integrations are 
performed by picking up the poles of the propagators $1/p_{(j)}^2$. Since 
the external field is smooth and $T_H$ is dominated by the hard scale, $n-1$ 
of the $n$ transverse momentum integrations can be performed trivially. This 
is not the case for the last integration which will necessarily be sensitive 
to the small off-shellness $k^2$. However, the $n-1$ performed integrations 
ensure that the eikonal factors associated with the high-$p_\perp$ partons 
are evaluated at the same transverse position. The resulting colour 
structure of the amplitude, after projection on a colour singlet final 
state, involves the trace of $W_{x_\perp}(y_\perp)$ (cf. Eq.~\ref{wdef}). 
It is intuitively clear that only two eikonal factors are present since the 
high-$p_\perp$ partons are close together in transverse space. They are 
colour rotated like a single parton. 

Under the further assumption that final state momenta of the high-$p_\perp$ 
partons are not resolved on the soft scale, the following result is derived,
\be
\frac{d\sigma}{dX^{(n+1)}}=\frac{k_0^2\,|T_H|^2}{\pi\,q_0\,N_c}\,
\int_{x_\perp}\,\left|\,\int_{k_\perp}\frac{\mbox{tr}[\tilde{W}_{x_\perp}
(k_\perp'\!\!\!-\!k_\perp)]}{k^2}\,\right|^2\delta^2\!\left(
\mbox{$\sum$}p_{(j)\perp}\right)\,\delta(q_0\!\!-\!q_0')\,.\label{cs2}
\ee
In this expression the non-perturbative input encoded in the Fourier 
transform $\tilde{W}$ is totally decoupled from the hard momenta that 
dominate $T_H$. 

The squared amplitude $|T_H|^2$ in Eq.~(\ref{cs2}) can be expressed through 
the partonic cross section $\hat{\sigma}(x,Q^2,y)$. In Fig.~\ref{fig:fac} 
this corresponds to the interpretation of the line labelled by $k$ as an 
incoming line for the hard process. The cross section differential in 
$\xi$ takes the form 
\be
\frac{d\sigma}{d\xi}=\int_x^\xi dy\,\hat{\sigma}(x,Q^2,y)\left(
\frac{df_s(y,\xi)}{d\xi}\right)\,.
\ee
Introducing the variable $b\!=\!y/\xi$ the diffractive parton density for 
scalars (integrated over $t$) reads 
\be
\frac{df_s(y,\xi)}{d\xi}\,=\frac{1}{\xi^2}\left(\frac{b}
{1-b}\right)\int\frac{d^2k_\perp'(k_\perp'^2)^2}{(2\pi)^4N_c}\int_{x_\perp}
\left|\int\frac{d^2k_\perp}{(2\pi)^2}\,\frac{\mbox{tr}[\tilde{W}_{x_\perp}
(k_\perp'\!\!\!-\!k_\perp)]}{k_\perp'^2b+k_\perp^2(1-b)}\right|^2\,.
\ee
Analogous considerations with spinor and vector partons result in the same 
factorizing result, but with different distribution functions. 

\begin{figure*}[t]
\begin{center}
\vspace*{-.1cm}
\parbox[b]{11cm}{\psfig{width=11cm,file=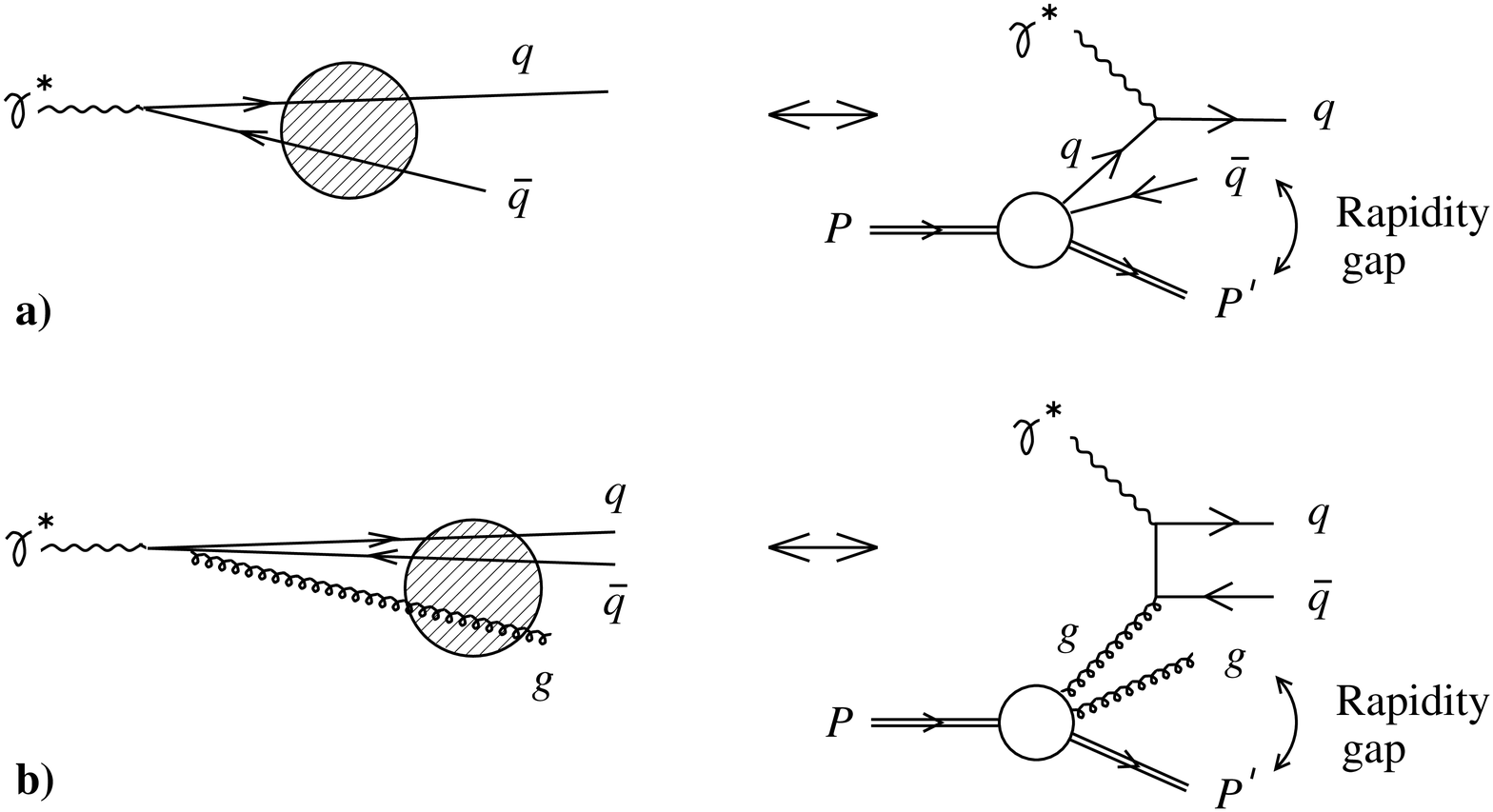}}\\
\vspace{-.4cm}
\end{center}
\caption{Diffractive DIS in the proton rest frame (left) and 
the Breit frame (right); asymmetric quark fluctuations correspond to 
diffractive quark scattering, asymmetric gluon fluctuations to diffractive 
boson-gluon fusion.}
\label{f2d}
\end{figure*}

The physical picture emerging from the above semiclassical calculation 
of diffractive parton distributions is illustrated in Fig.~\ref{f2d}. 
In the Breit frame, leading order diffractive DIS is most naturally 
described by photon-quark scattering, with the quark coming from the 
diffractive parton distribution of the target hadron. This is illustrated 
on the r.h. side of Fig.~\ref{f2d}a. Identifying the leading twist part 
of the $q\bar{q}$ pair production cross section (l.h. side of 
Fig.~\ref{f2d}a) with the result of the conventional partonic calculation 
(r.h. side of Fig.~\ref{f2d}a), the diffractive quark distribution of 
the target is expressed in terms of the color field dependent function 
given in Eq.~(\ref{wdef}). 

Similarly, the cross section for the color singlet production of a 
$q\bar{q}g$ state (l.h. side of Fig.~\ref{f2d}b) is identified with the 
boson-gluon fusion process based on the diffractive gluon distribution of 
the target (r.h. side of Fig.~\ref{f2d}b). This allows for the calculation 
of the diffractive gluon distribution in terms of a function similar to 
Eq.~(\ref{wdef}) but with the $U$ matrices in the adjoint representation. 

In the semiclassical approach, the cross sections for inclusive DIS are 
obtained from the same calculations as in the diffractive case where, 
however, the color singlet condition for the final state parton 
configuration is dropped. As a result, the $q\bar{q}$ production cross 
section (cf. the l.h. side of Fig.~\ref{f2d}a) receives contributions from 
both the aligned jet and the high-$p_\perp$ region. In the latter, the 
logarithmic $d p_\perp^2/p_\perp^2$ integration gives rise to a $\ln Q^2$ 
term in the full cross section. 

\begin{figure}[t]
\begin{center}
\vspace*{.2cm}
\parbox[b]{11.3cm}{\psfig{width=11.3cm,file=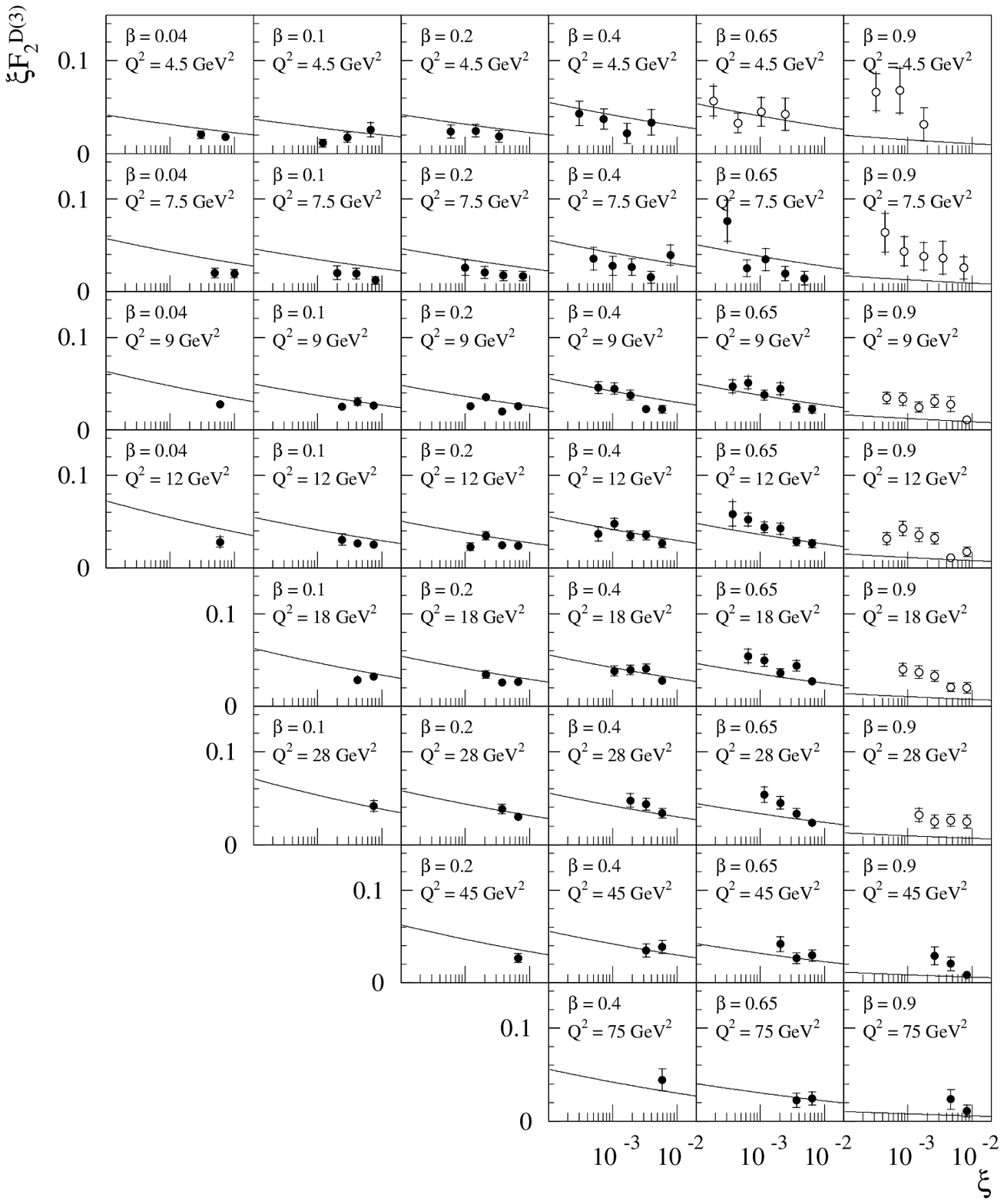}}
\end{center}\vspace*{-.3cm}
\refstepcounter{figure}
\label{h1}
{\bf Figure \ref{h1}:} 
The structure function $F_2^{D(3)}(\xi,\beta,Q^2)$ computed in the 
semiclassical approach with H1 data from~\protect\cite{nh1}. The open data 
points correspond to $M^2 \leq 4$~GeV$^2$ and are not included in the fit. 
\end{figure}

\begin{figure}[t]
\vspace*{-3.2cm}
\begin{center}
\parbox[b]{5cm}{\psfig{file=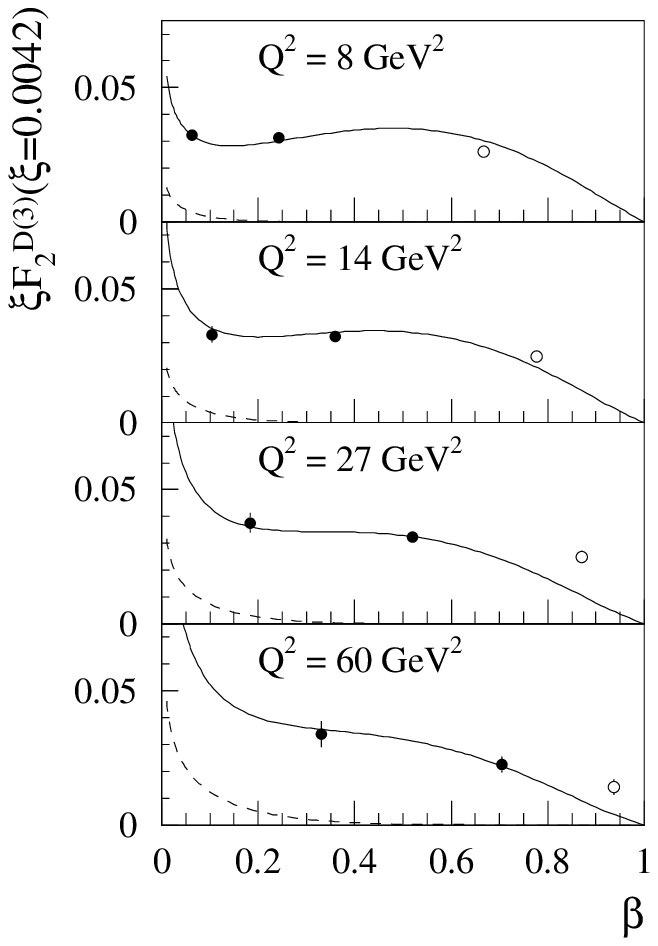,width=5cm}}\\
\vspace{-3.4cm}
\end{center}
\refstepcounter{figure}
\label{zeus}
{\bf Figure \ref{zeus}:} The diffractive structure function 
$F_2^{D(3)}(\xi,\beta,Q^2)$ with data from ZEUS~\cite{nzeus}. Open circles 
correspond to $M^2\leq 4$~GeV$^2$. The charm content is indicated as a 
dashed line.
\vspace*{-.2cm}
\end{figure}

In the leading order partonic analysis, the full cross section is described 
by photon-quark scattering. The gluon distribution is responsible for the 
scaling violations at small $x$, $\partial F_2(x,Q^2)/\partial\ln Q^2 
\sim xg(x,Q^2)$. Thus, the semiclassical result for $q\bar{q}$ production, 
with its $\ln Q^2$ contribution, is sufficient to calculate both the 
inclusive quark and the inclusive gluon distribution. The results are again 
expressed in terms of the function in Eq.~(\ref{wdef}) where now the color 
trace is taken {\it after} the two $W$ matrices (corresponding to the 
amplitude and its complex conjugate) have been multiplied. 

To obtain explicit formulae for the above parton distributions, a model for 
the averaging over the color fields has to be introduced. Such models are 
provided, e.g., by the non-perturbative stochastic vacuum~\cite{dos} or by 
the perturbative small dipoles of~\cite{hks}. The large hadron 
model~\cite{mv}, on which the following analysis is based, has the the 
advantage of justifying the Fock space expansion of the photon wave 
function while being intrinsically non-perturbative as far as the $t$ 
channel colour exchange is concerned. 

In the case of a very large hadronic target~\cite{mv} (see 
also~\cite{mcl}), even in the aligned 
jet region, the transverse separation of the $q\bar{q}$ pair remains 
small~\cite{hw}. This is a result of the 
saturation of the dipole cross section at smaller dipole size. Under the 
additional assumption that color fields in distant regions of the large 
target are uncorrelated, a simple Glauber-type exponentiation of the 
averaged local field strength results in explicit formulae for all the 
relevant parton distribution functions~\cite{bgh} (see~\cite{kov} for 
a closely related analysis). 

Thus, diffractive and inclusive quark and gluon distributions at some 
small scale $Q_0^2$ are expressed in terms of only two parameters, the 
average color field strength and the total size of the large target 
hadron. The energy dependence arising from the large-momentum cutoff 
applied in the process of color field averaging can not be calculated from 
first principles. It is described by a $\ln^2 x$ ansatz, consistent with 
unitarity, which is universal for both the inclusive and diffractive 
structure function~\cite{b}. This introduces a further parameter, the 
unknown constant that comes with the logarithm (see~\cite{gbw1} for a 
related but different way of introducing a phenomenological energy 
dependence in diffractive and inclusive DIS). 

A conventional leading order DGLAP analysis of data at small $x$ and 
$Q^2>Q_0^2$ results in a good four parameter fit ($Q_0$ being the fourth 
parameter) to both the inclusive and diffractive structure function 
(Figs.~\ref{h1} and \ref{zeus}). Diffractive data with $M^2<4\,\mbox{GeV}^2$ 
is excluded from the fit since higher twist effects are expected to affect 
this region (cf.~\cite{bekw} for a phenomenological discussion of higher 
twist effects). As an illustration, the $\beta$ dependence of $F_2^{D(3)}$ 
at different values of $Q^2$ is shown in Figs.~\ref{h1} and \ref{zeus} 
(see~\cite{bgh} for further plots, in particular of the inclusive structure 
function, and more details of the analysis). 

Two important qualitative features of the approach should be 
emphasized. First, the diffractive gluon distribution is much larger than 
the diffractive quark distribution, a result reflected in the pattern of 
scaling violations of $F_2^{D(3)}$. This feature is also present in the 
analysis of~\cite{hks}, where, in contrast to the present approach, the 
target is modelled as a small color dipole. Second, the inclusive gluon 
distribution, calculated from $q\bar{q}$ pair production at high $p_\perp$ 
and determined by the small-distance structure of the color field, is large 
and leads to the dominance of inclusive over diffractive DIS.

\section{Vector meson production}\label{vm}
So far, inclusive diffraction, as parametrized, e.g., by the diffractive 
structure function $F_2^D$, was at the centre of interest of this review. 
It was argued that, for inclusive processes, the underlying colour singlet 
exchange is soft. In perturbative QCD, the simplest possibility of 
realizing colour singlet exchange is via two $t$ channel gluons. In fact, 
the colour singlet exchange in certain more exclusive diffractive processes 
is, with varying degree of rigour, argued to be governed by a hard scale. 
In such cases, two-gluon exchange dominates.

\subsection{Elastic meson production}\label{sect:emp}
Elastic meson electroproduction is the first diffractive process that was 
claimed to be calculable in perturbative QCD~\cite{rys,bro}. It has since 
been considered by many authors, and a fair degree of understanding has been 
achieved as far as the perturbative calculability and the factorization of 
the relevant non-perturbative parton distributions and meson wave functions 
are concerned. 

\begin{figure}[ht]
\begin{center}
\vspace*{.2cm}
\parbox[b]{7.5cm}{\psfig{width=7.5cm,file=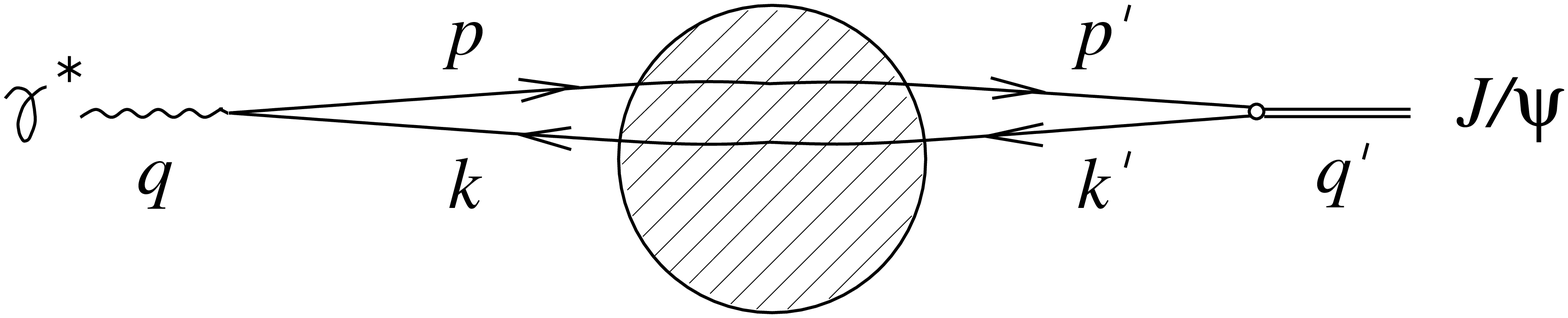}}
\end{center}
\refstepcounter{figure}
\label{fig:jpsi}
{\bf Figure \ref{fig:jpsi}:} Leading order amplitude for the elastic 
production of a $J/\Psi$ particle off an external colour field. 
\end{figure}

To begin with, consider the electroproduction of a heavy $q\bar{q}$ bound 
state off a given classical colour field. The relevant amplitude is shown 
in Fig.~\ref{fig:jpsi}. In the non-relativistic limit, the two outgoing 
quarks are on-shell, and each carries half of the $J/\psi$ momentum. 
Thus, the two quark propagators with momenta $p'=k'=q'/2$ and the $J/\psi$ 
vertex are replaced with the projection operator $g_J\epsilons_J(\ks'+m)$. 
Here 
\be
g_J^2=\frac{3\Gamma^J_{ee}m_J}{64\pi\aem^2}\,,
\ee
$\Gamma^J_{ee}$ is the electronic decay width of the $J/\psi$ particle, 
$m_J=2m$ is its mass, and $\epsilon_J$ its polarization vector~\cite{bj}. 

Using the calculational technique of Sect.~\ref{sa}, the amplitude of 
Fig.~\ref{fig:jpsi} can be expressed in terms of non-Abelian eikonal 
factors $U$ and $U^\dagger$ associated with the two quarks. 

Considering both $Q$ and $m$ as hard scales, while $U$ and $U^\dagger$ are 
governed by the soft hadronic scale $\Lambda$, the integrand in the loop 
integration of Fig.~\ref{fig:jpsi} can be expanded in powers of the soft 
momentum $k_\perp$. The leading power of the amplitude is given by the 
first non-vanishing term. In the case of forward production, $p_\perp'= 
k_\perp'=0$, the dependence on the external colour field takes the 
form~\cite{hab}
\bea
&&\int d^2k_\perp k_\perp^2 \mbox{tr}\left[\tilde{U}(p_\perp'\!-\!p_\perp)
\tilde{U}^\dagger(k_\perp\!-\!k_\perp')-(2\pi)^4\delta^2(p_\perp'\!-\!
p_\perp)\delta^2(k_\perp'\!-\!k_\perp)\right]\nonumber
\\
&&\hspace*{2cm}=\,-(2\pi)^2\partial_{y_\perp}^2\int_{x_\perp}\mbox{tr}
W_{x_\perp}(y_\perp)\Big|_{y_\perp=0}\,\,.\label{ddw}
\eea

\begin{figure}[ht]
\begin{center}
\vspace*{.2cm}
\parbox[b]{7.5cm}{\psfig{width=7.5cm,file=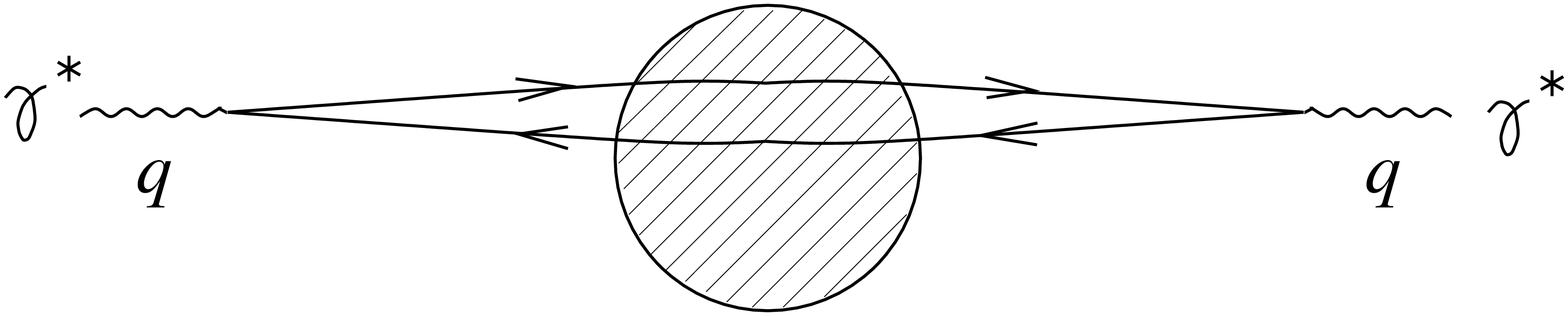}}
\end{center}
\refstepcounter{figure}
\label{fig:comp}
{\bf Figure \ref{fig:comp}:} The Compton scattering amplitude within the 
semiclassical approach. 
\end{figure}

Now, the crucial observation is that precisely the same dependence on the 
external field is present in the amplitude for forward Compton scattering 
shown in Fig.~\ref{fig:comp}. In the case of longitudinal photon 
polarization, the transverse size of the $q\bar{q}$ pair is always small, 
and the target field enters only via the second derivative of $W$ that 
appears in Eq.~(\ref{ddw}). 

Thus, comparing with the parton model result for longitudinal photon 
scattering, this derivative can be identified in terms of the gluon 
distribution of the target proton~\cite{bhmpt}, 
\be
-\partial_{y_\perp}^2\int_{x_\perp}\mbox{tr}W_{x_\perp}(y_\perp)
\Big|_{y_\perp=0}=2\pi^2\alpha_sxg(x)\,.\label{gd}
\ee
Using this relation, the amplitudes for the forward production of 
transversely and longitudinally polarized $J/\psi$ mesons by transversely 
and longitudinally polarized virtual photons are obtained. Under the 
additional assumption $Q^2\gg m_J^2$, the amplitudes for longitudinal and 
transverse polarization read 
\be
T_L=-i64\pi^2\alpha_sg_Je\,(xg(x))\frac{s}{3Q^3}\quad,\quad T_T=
\frac{m_J}{Q}T_L\,,\label{tltt}
\ee
where $\sqrt{s}$ is the centre-of-mass energy of the $\gamma^*p$ 
collision. This is Ryskin's celebrated result for elastic $J/\psi$ 
production~\cite{rys}. 

It is not surprising that the gluon distribution of Eq.~(\ref{gd}), 
calculated according to Fig.~\ref{fig:comp}, shows no scaling violations and 
only the trivial Bremsstrahlungs energy dependence $\sim 1/x$. The reason 
for this is the softness assumptions of the semiclassical calculation. 
Firstly, the eikonal approximation implies that all longitudinal modes of 
the external field are much softer than the photon energy. Secondly, the 
reduction of the field dependence to a transverse derivative is only 
justified if the scales governing the quark loop, i.e., $Q^2$ in the case 
of Fig.~\ref{fig:comp} and $Q^2$ and $m^2$ in the case of 
Fig.~\ref{fig:jpsi}, are harder than the transverse structure of $W$. 
These two approximations, evidently valid for a given soft field, are also 
justified for a dynamical target governed by QCD as long as only leading 
logarithmic accuracy in both $1/x$ and $Q^2$ is required. Thus, a 
non-trivial dependence on $1/x$ and $Q^2$ can be reintroduced into 
Eq.~(\ref{tltt}) via the measured gluon distribution, keeping in mind that 
the result is only valid at double-leading-log accuracy. 

\begin{figure}[ht]
\begin{center}
\vspace*{.2cm}
\parbox[b]{4cm}{\psfig{width=4cm,file=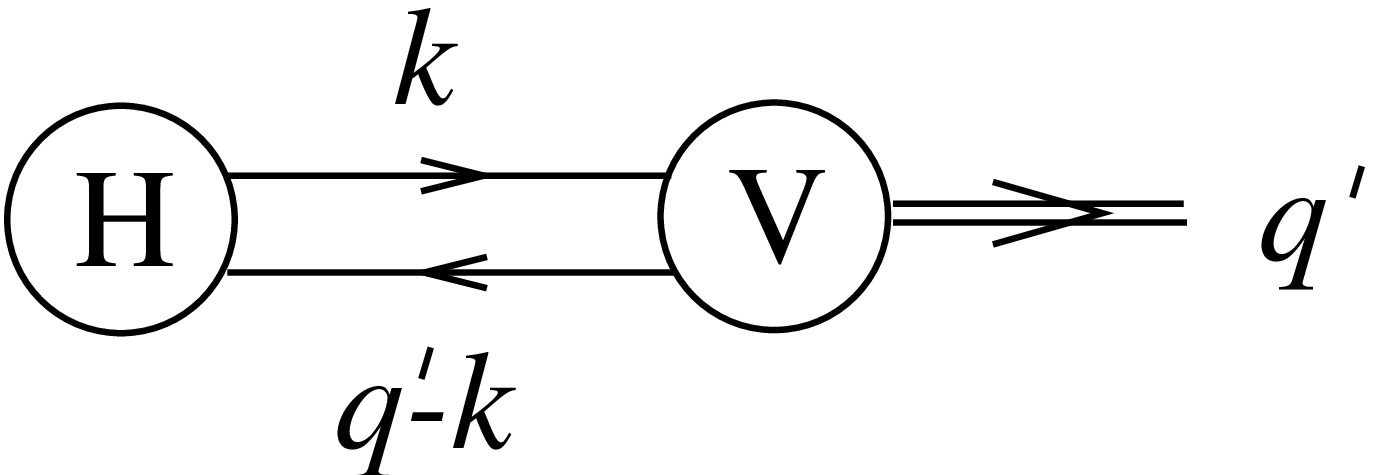}}
\end{center}
\refstepcounter{figure}
\label{fig:hs}
{\bf Figure \ref{fig:hs}:} Generic diagram for meson production in a hard 
process. 
\end{figure}

An essential extension of the above fundamental result is related to 
the treatment of the bound state produced. Brodsky et al.~\cite{bro} showed 
that, at least for longitudinal photon polarization, a perturbative 
calculation is still possible in the case of light, non-perturbative bound 
states like the $\rho$ meson. The calculation is based on the concept of 
the light-cone wave function of this meson (see, however,~\cite{mrt} for an 
alternative approach to light meson production). Referring the reader 
to~\cite{cz} for a detailed review, a brief description of the main ideas 
involved is given below (cf.~\cite{pire}). For this purpose, consider the 
generic diagram for the production of a light meson in a hard QCD process 
given in Fig.~\ref{fig:hs}. 

Assume that, as shown in this figure, all diagrams can be cut across two 
quark-lines, the constituent quarks of the meson, in such a way as to 
separate the hard process H from the soft meson formation vertex V, which 
is defined to include the propagators. The amplitude can be written as 
\be
T\!=\!\int\!d^4k\,T_H(k)V(k)\!=\!\int_0^1\!dz\,T_H(z)\,\frac{q'_+}{2}\!\int
\!dk_-d^2k_\perp V(k)\!=\!\int_0^1\!dz\,T_H(z)\phi(z), 
\ee
where $z=k_+/q_+'$, and the last equality is simply the definition of the 
light-cone wave function $\phi$ of the meson. The two crucial 
observations leading to the first of these equalities are the approximate 
$k_-$ and $k_\perp$ independence of $T_H$ and the restriction of the $z$ 
integration to the interval from 0 to 1. The first is the result of the hard 
scale that dominates $T_H$, the second follows from the analytic structure 
of $V$. In QCD, the $k_\perp$ integration implicit in $\varphi$ usually 
has an UV divergence due to gluon exchange between the quarks. Therefore, 
one should really read $\phi=\phi(z,\mu^2)$, where the cutoff $\mu^2$ is of 
the order of the hard scale that governs $T_H$. At higher orders in 
$\alpha_s$, the hard amplitude $T_H$ develops a matching IR cutoff 
dependence. 

The discussion of vector meson production given so far was limited to the 
double-leading-log approximation as far as the colour singlet exchange in 
the $t$ channel is concerned. To go beyond this approximation, the concept 
of `non-forward' or `off-diagonal' parton distributions, introduced some 
time ago (see~\cite{mea} and refs. therein) and discussed by Ji~\cite{ji} 
and Radyushkin~\cite{rad} in the present context, has to be used
(see also~\cite{gm}).

\begin{figure}[ht]
\begin{center}
\vspace*{.2cm}
\parbox[b]{7.5cm}{\psfig{width=7.5cm,file=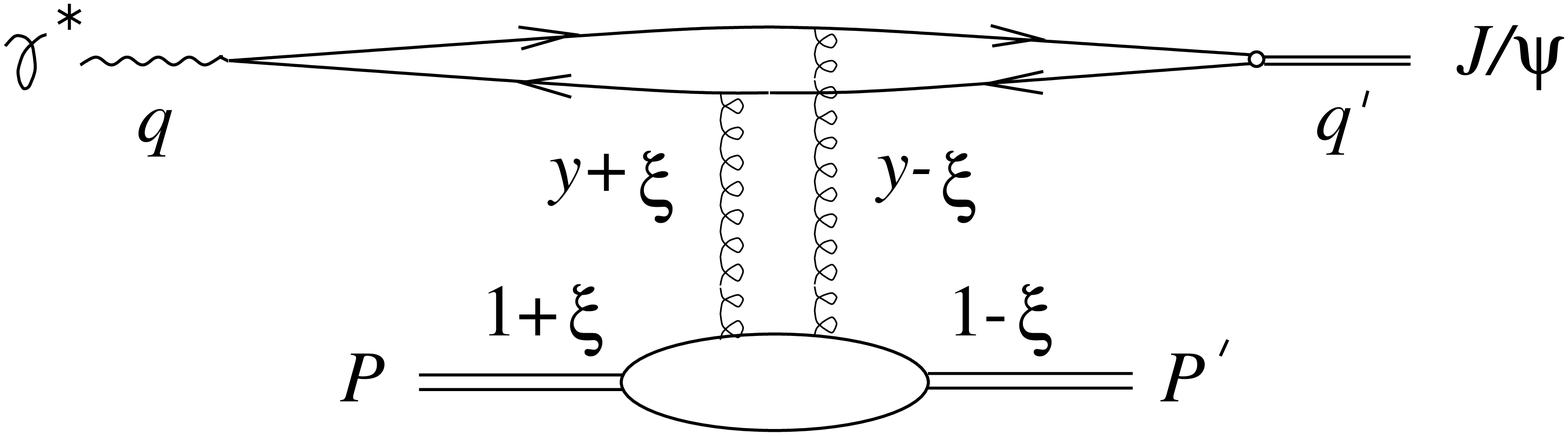}}
\end{center}
\refstepcounter{figure}
\label{fig:jpsi1}
{\bf Figure \ref{fig:jpsi1}:} Elastic $J/\psi$ production (three further 
diagrams with the gluons connected in different ways have to be added). The 
two gluon lines and the incoming and outgoing proton are labelled by their 
respective fractions of the plus component of $\bar{P}\equiv(P+P')/2$. 
\end{figure}

Recall first that the semiclassical viewpoint of Figs.~\ref{fig:jpsi} and 
\ref{fig:comp} is equivalent to two-gluon exchange as long as the 
transverse size of the energetic $q\bar{q}$ state is small. So far, the 
recoil of the target in longitudinal direction has been neglected. However, 
such a recoil is evidently required by the kinematics. For what follows, it 
is convenient to use a frame where $q_-$ is the large component of the 
photon momentum. In Fig.~\ref{fig:jpsi1}, the exchanged gluons and the 
incoming and outgoing proton with momenta $P$ and $P'$ are labelled by their 
respective fractions of the plus component of $\bar{P}\equiv(P+P')/2$. If 
$\Delta$ is the momentum transferred by the proton, $\xi\bar{P}_+=\Delta_+ 
/2$. The variable $y$ is an integration variable in the gluon loop. 

The lower part of the diagram in Fig.~\ref{fig:jpsi1} is a generalization 
of the conventional gluon distribution. It can be described by the 
non-forward gluon distribution 
\be
H_g(y,\xi,t)=\frac{1}{4\pi y\bar{P}_+}\int dx_- e^{-iy\bar{P}_+x_-/2}\langle 
P'|F^\dagger(0,x_-,0_\perp)^{+\mu}F(0,0,0_\perp)_\mu{}^+|P\rangle\,.
\label{nfpd}
\ee

The description of elastic meson production in terms of non-forward parton 
distributions is superior to the double-leading-log approach $\!\!\mbox{
of~\cite{rys,bro}}$ since $\alpha_s$ corrections to the hard amplitude, meson 
wave function and parton distribution function can, at least in principle, 
be systematically calculated. However, the direct relation to the measured 
conventional gluon distribution is lost. A new non-perturbative quantity, 
the non-forward gluon distribution, is introduced, which has to be measured 
and the evolution of which has to be tested -- a very complicated problem 
given the uncertainties of the experiment and of the meson wave functions 
involved (see~\cite{fnf} for possibilities of predicting the non-forward 
from the forward distribution functions).

\subsection{Factorization}\label{sect:fact}
Having discussed the leading order results for vector meson production, 
the next logical step is to ask whether the systematic calculation of 
higher order corrections is feasible. For this, it is necessary to 
understand the factorization properties of the hard amplitude and the two 
non-perturbative objects involved, i.e., the meson wave function and the 
non-forward gluon distribution. 

Factorization means that, to leading order in $1/Q$, the amplitude 
can be written as 
\be
T=\int_0^1dz\int dy\,H(y,x/2)\,T_H(Q^2,y/x,z,\mu^2)\,\phi(z,\mu^2)\,, 
\label{vma}
\ee
where $T_H$ is the hard scattering amplitude, $\phi$ is the light-cone 
wave function of the vector meson produced, and $H$ is the non-forward 
parton distribution of the proton. It could, for example, be the 
non-forward gluon distribution $H_g$ of the last section. The variable 
$x=\xbj$ is the usual DIS Bjorken variable. 

\begin{figure}[ht]
\begin{center}
\vspace*{.2cm}
\parbox[b]{11cm}{\psfig{width=11cm,file=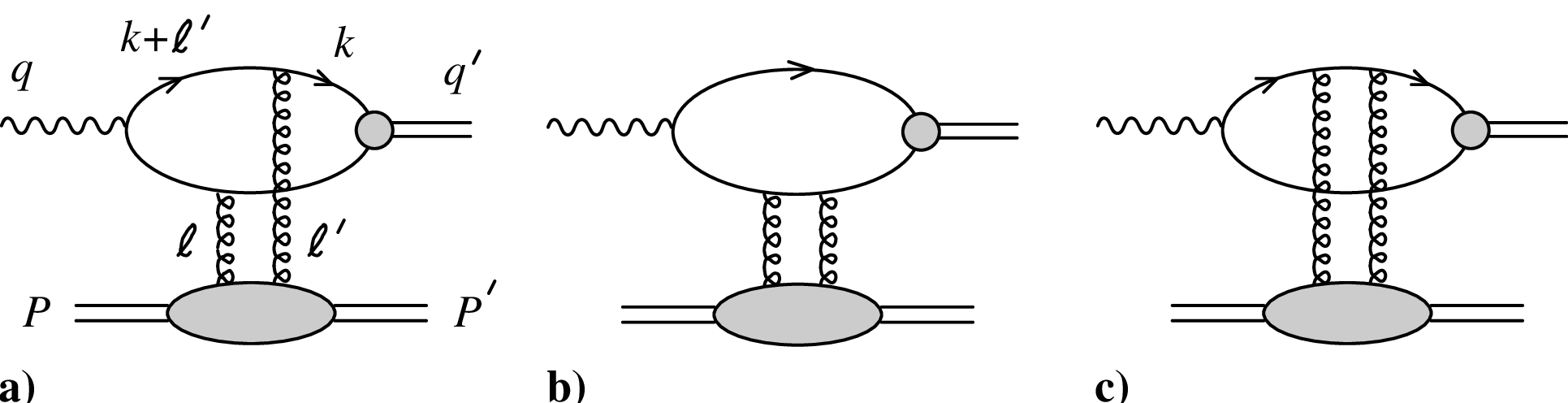}}
\end{center}
\refstepcounter{figure}
\label{fig:lo}
{\bf Figure \ref{fig:lo}:} The leading amplitude for a point-like meson 
vertex. 
\end{figure}

In these notes, I will not describe the factorization proof for 
longitudinal vector meson production given in~\cite{cfs1}. Instead, a 
particularly simple situation will be used to show how factorization works 
specifically at small $x$, from the point of view of the target rest 
frame~\cite{hl}. For this purpose, consider a very energetic scalar photon 
that scatters off a hadronic target producing a scalar meson built from two 
scalar quarks (see Fig.~\ref{fig:lo}). The quarks are coupled to the photon 
and the meson by point-like scalar vertices $ie$ and $i\lambda$, where $e$ 
and $\lambda$ have dimension of mass. The coupling of the gluons to the 
scalar quarks is given by $-ig\,(r_\mu+r_\mu')$, where $r$ and $r'$ are the 
momenta of the directed quark lines, and $g$ is the strong gauge coupling. 

Under quite general conditions \cite{hl}, the gluon momenta satisfy the 
relations $\ell_+,\,\ell_+'\ll q_+,$ $\ell_-,\,\ell_-'\ll P_-$ and $\ell^2 
\sim\ell^{\prime 2} \sim-\ell_\perp^2$. Then the lower bubble in 
Fig.~\ref{fig:lo} effectively has the structure 
\be
F^{\mu\nu}(\ell,\ell',P)\simeq\delta(P_-\ell_+)\,F(\ell_\perp^2)\,P^\mu 
P^\nu\,,\label{fd}
\ee
which is defined to include both gluon propagators and all colour factors. 

Assume that $F$ restricts the gluon momentum to be soft, $\ell_\perp^2\ll 
Q^2$. In the high-energy limit, it suffices to calculate 
\be
M=\int\frac{d^4\ell}{(2\pi)^4}T^{\mu\nu}F_{\mu\nu}\simeq\int
\frac{d^4\ell}{4(2\pi)^4}T_{++}F_{--}\,,\label{it}
\ee
where 
\be
T^{\mu\nu}=T^{\mu\nu}(\ell,\ell',q)=T^{\mu\nu}_a+T^{\mu\nu}_b+T^{\mu\nu}_c
\label{tmn}
\ee
is the sum of the upper parts of the diagrams in Fig.~\ref{fig:lo}. 

Note that, because of the symmetry of $F_{\mu\nu}$ with respect to the two 
gluon lines, the amplitude $T^{\mu\nu}$ of Eq.~(\ref{tmn}) is used instead 
of the properly-symmetrized upper amplitude 
\be
T_{\mbox{\scriptsize sym}}^{\mu\nu}(\ell,\ell',q)=\frac{1}{2}[T^{\mu\nu}
(\ell,\ell',q)+ T^{\nu\mu}(-\ell',-\ell,q)]\,.
\ee
The two exchanged gluons together form a colour singlet and so the 
symmetrized amplitude $T_{\mbox{\scriptsize sym}}^{\mu\nu}$ satisfies the 
same Ward identity as for two photons, 
\be
T_{\mbox{\scriptsize sym}}^{\mu\nu}(\ell,\ell',q)\ell_\mu \ell_\nu'=0\,.
\label{w1}
\ee
Writing this equation in light-cone components and setting $\ell_\perp=
\ell_\perp'$, as appropriate for forward production, it follows that, for 
the relevant small values of $\ell _-$, $\ell '_-$, $\ell_+$ and $\ell_+'$, 
\be
T_{\mbox{\scriptsize sym,}++} \sim\ell_\perp^2\,
\ee
in the limit $\ell_\perp^2\to 0$. Here the fact that the tensor 
$T_{\mbox{\scriptsize sym}}^{\mu\nu}$, which is built from $\ell'$, $\ell$ 
and $q$, has no large minus components has been used. The $\ell_-$ 
integration makes this equation hold also for the original, unsymmetrized 
amplitude, 
\be
\int d\ell_-T_{++} \sim\ell_\perp^2\,.\label{w2}
\ee
This is the crucial feature of the two-gluon amplitude that will simplify 
the calculation and lead to the factorizing result below. 

Consider first the contribution from diagram a) of Fig.~\ref{fig:lo} to 
the $\ell_-$ integral of $T_{++}$, which is required in Eq.~(\ref{it}), 
\be
\int d\ell_-T_{a,++}=-4eg^2q_+\int\frac{d^4k}{(2\pi)^3}\,\frac{z(1-z)}
{N^2+(k_\perp+\ell_\perp)^2}\,\frac{i\lambda}{k^2(q'-k)^2}\,.
\label{ta}
\ee
Here $N^2=z(1-z)Q^2$, $z=k_+/q_+$ and the condition $\ell_+=0$, enforced 
by the $\delta$-function in Eq.~(\ref{fd}), has been anticipated. 

Now $\int d\ell_-T_{b,++}$ and $\int d\ell_-T_{c,++}$ each carry no 
$\ell_\perp$ dependence. So, to ensure the validity of Eq.~(\ref{w2}), the 
sum of the three diagrams must be 
\be
\int d\ell_-T_{++}=4eg^2q_+\int\frac{d^4k}{(2\pi)^3}z(1-z){\cal N}
\frac{i\lambda}{k^2(q'-k)^2}\,,
\label{intit}
\ee
where
\be
{\cal N}=\Big [\frac{1}{N^2+k_\perp^2}-\frac{1}
{N^2+(k_\perp+\ell_\perp)^2}\Big ]\sim \frac{\ell_\perp^2}
{(N^2+k_\perp^2)^2}\,.\label{intitt}
\ee
Note the $1/Q^4$ behaviour obtained after a cancellation of $1/Q^2$ 
contributions from the individual diagrams. This cancellation is closely 
related to the well-known effect of colour transparency.

Introduce the $k_\perp$ dependent light-cone wave function of the meson 
\be
\phi(z,k_\perp^2)=-\frac{iq'_+}{2}\int dk_-dk_+\,
\frac{i\lambda}{(2\pi)^4k^2(q'-k)^2}\,\delta (k_+-zq'_+).\label{wv}
\ee
The final result following from Eqs.~(\ref{it}) and (\ref{intit}) is a 
convolution of the production amplitude of two on-shell quarks and the 
light-cone wave function:
\be
M=ieg^2s\left(\int\frac{d^2\ell_\perp}{2(2\pi)^3}\ell_\perp^2F(
\ell_\perp^2)\right)\int dz\int d^2k_\perp\frac{z(1-z)}{(N^2+k_\perp^2)^2}
\phi(z,k_\perp^2)\,.\label{lot}
\ee
This corresponds to the $O(\ell_\perp^2)$ term in the Taylor expansion
of the contribution from Fig.~\ref{fig:lo}a, given in Eq.~(\ref{ta}). 

\begin{figure}[ht]
\begin{center}
\vspace*{.2cm}
\parbox[b]{4cm}{\psfig{width=4cm,file=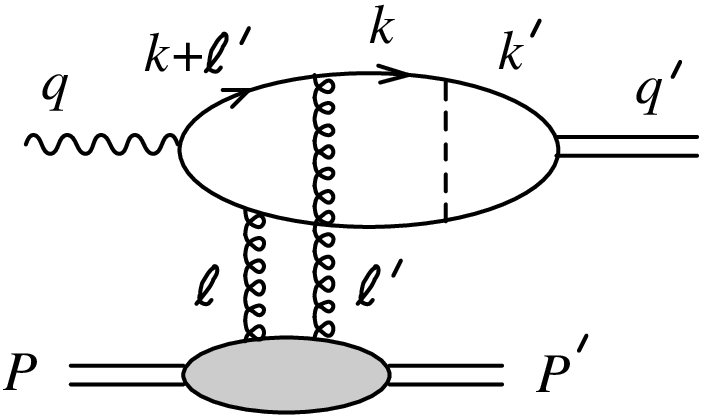}}
\end{center}
\refstepcounter{figure}
\label{fig:nlo}
{\bf Figure \ref{fig:nlo}:} Diagram for meson production with the vertex 
modelled by scalar particle exchange.
\end{figure}

At leading order, factorization of the meson wave function was trivial 
since the point-like quark-quark-meson vertex $V(k^2,(q'-k)^2)=i\lambda$ 
was necessarily located to the right of the all other interactions. To 
see how factorization comes about in the simplest non-trivial situation, 
consider the vector meson vertex 
\be
V(k^2,(q'-k)^2)=\int\frac{d^4k'}{(2\pi)^4}\,\frac{i\lambda\lambda'^2}{k'^2
(q'-k')^2(k-k')^2}\,,\label{tf2v}
\ee
which corresponds to the triangle on the r.h. side of Fig.~\ref{fig:nlo}. 
Here, the dashed line denotes a colourless scalar coupled to the scalar 
quarks with coupling strength $\lambda'$. 

\begin{figure}[ht]
\begin{center}
\vspace*{-.2cm}
\parbox[b]{11cm}{\psfig{width=11cm,file=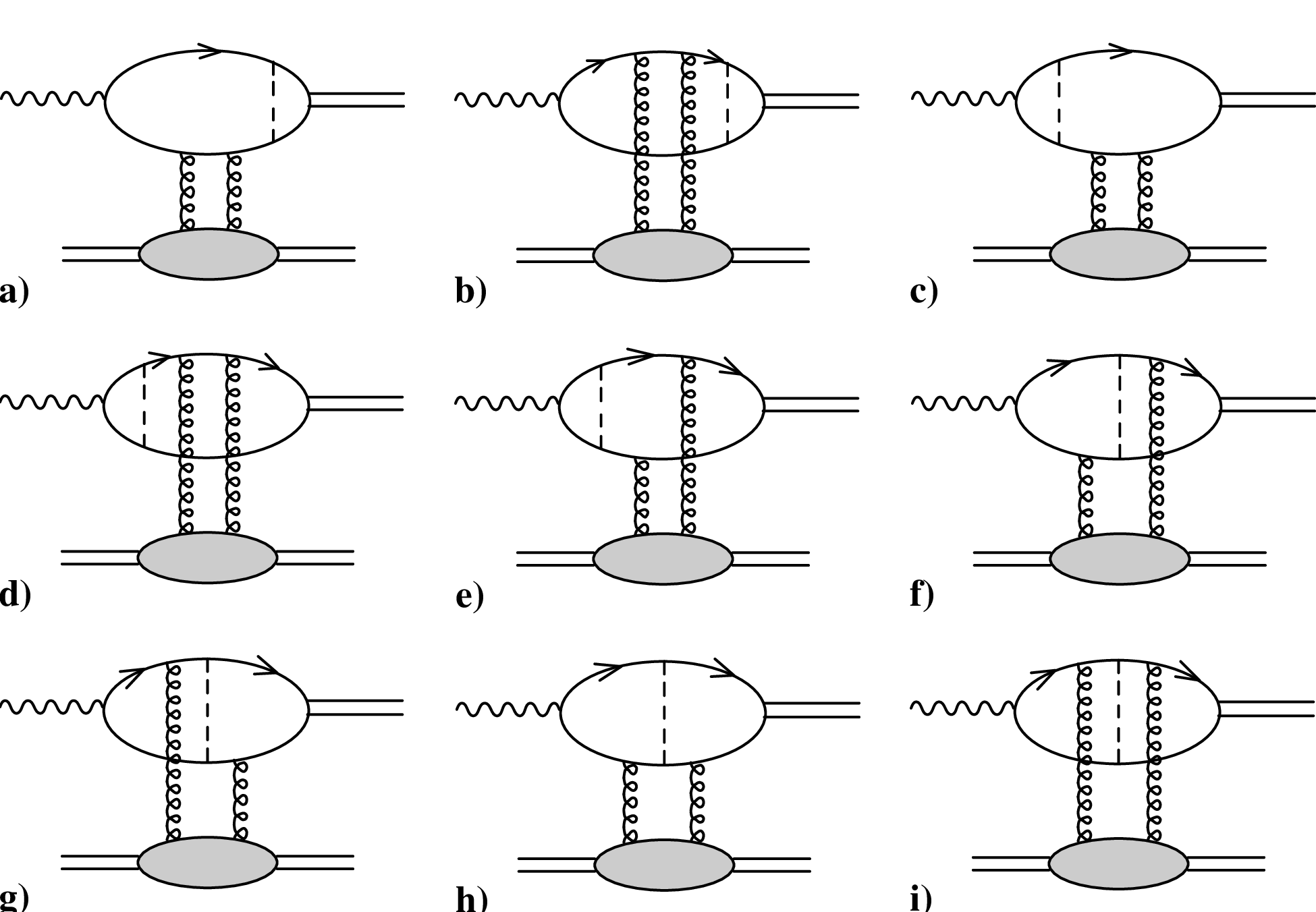}}
\end{center}
\refstepcounter{figure}
\label{fig:rest}
{\bf Figure \ref{fig:rest}:} The remaining diagrams contributing to meson 
production within the above simple model for the meson wave function.
\end{figure}

The diagram of Fig.~\ref{fig:nlo} by itself gives no consistent description 
of meson production since it lacks gauge invariance. This problem is not 
cured by just adding the two diagrams \ref{fig:lo}b) and c) with the blob 
replaced by the vertex $V$. It is necessary to include all the diagrams 
shown in Fig.~\ref{fig:rest}.

The same gauge invariance arguments that lead to Eq.~(\ref{w2}) apply to 
the sum of all the diagrams in Figs.~\ref{fig:nlo} and \ref{fig:rest}. 
Therefore, the complete result for $T_{++}$, which is now defined by the 
sum of the upper parts of all these diagrams, can be obtained by extracting 
the $\ell_\perp^2$ term at leading order in the energy and $Q^2$. Such a 
term, with a power behaviour $\sim \ell_\perp^2/Q^4$, is obtained from 
the diagram in Fig.~\ref{fig:nlo} (replace $i\lambda$ in Eq.~(\ref{ta}) 
with the vertex $V$ of Eq.~(\ref{tf2v})) by expanding around $\ell_\perp=0$. 
It can be demonstrated that none of the other diagrams gives rise to such a 
leading-order $\ell_\perp^2$ contribution (see~\cite{hl} for more details). 

The complete answer is given by the $\ell_\perp^2$ term from the Taylor 
expansion of Eq.~(\ref{ta}). The amplitude $M$ is precisely the one of 
Eqs.~(\ref{lot}) and (\ref{wv}), with $i\lambda$ substituted by $V$ of 
Eq.~(\ref{tf2v}). The correctness of this simple factorizing result has also 
been checked by explicitly calculating all diagrams of Fig.~\ref{fig:rest}. 

The above simple model calculation can be summarized as follows. The 
complete result contains leading contributions from diagrams that cannot 
be factorized into quark-pair production and meson formation. However, the 
answer to the calculation can be anticipated by looking only at one 
particular factorizing diagram. The reason for this simplification is gauge 
invariance. In the dominant region, where the transverse momentum 
$\ell_\perp$ of the two $t$-channel gluons is small, gauge invariance 
requires the complete quark part of the amplitude to be proportional to 
$\ell_\perp^2$. The leading $\ell_\perp^2$ dependence comes exclusively 
from one diagram. Thus, the complete answer can be obtained from this 
particular diagram, which has the property of factorizing explicitly if the 
two quark lines are cut. The resulting amplitude can be written in a 
factorized form. 

Note finally that, although QCD predicts factorization for longitudinal 
vector meson production and a $1/Q^2$ suppression of the transverse cross 
section, this behaviour is not visible in the data~\cite{vd}. A recent 
calculation~\cite{cro} explaining the data is, in my opinion, in conflict 
with QCD expectations and, in particular, with the implications 
of~\cite{hl}, since it employs a non-local vector meson vertex in a 
loop calculation. It is important to clarify this situation.

\section{Summary}\label{sum}
The phenomenon of diffraction in DIS can be understood in very simple 
terms if the process is viewed in the rest frame of the target proton. 
The energetic virtual photon develops a partonic fluctuation which then 
scatters off the target. A certain fraction of the total DIS cross section 
is due to photon fluctuations with large transverse size. This fraction is 
not power suppressed in the high-$Q^2$ limit. Since, as one expects from 
the experience with hadron-hadron collisions, part of these large size 
fluctuations scatter quasi-elastically off the proton, a leading twist 
diffractive cross section is obtained. 

One possibility to describe the interaction of the photon fluctuation 
with the target is two-gluon exchange. In certain more exclusive 
processes, such as longitudinally polarized vector meson production, the 
transverse size of the relevant photon fluctuations is always small and
two-gluon calculations can be rigorously justified. However, for most of 
the diffractive cross section no suppression of multiple gluon exchange 
can be derived. 

This problem is addressed in the semiclassical approach, where the target 
is described by a superposition of soft colour fields and gluon exchange 
is resummed to all orders in an eikonal approximation. Diffraction arises 
whenever the partonic fluctuation of the photon remains in a colour 
singlet. The application of the semiclassical approach to experimental 
results is particularly simple if the approach is used to derive both 
diffractive and inclusive parton distributions at some small input scale. 
In this case, the analysis of all higher-$Q^2$ data proceeds with standard 
perturbative methods. Different models for the underlying colour fields 
can be compared to diffractive and inclusive structure function data in a 
very direct way.\\[.4cm]
{\bf Acknowledgements}\\[.1cm]
I have greatly benefited from my work with W. Buchm\"uller, M.F. McDermott 
and T. Gehrmann, the results of which are reflected in part of these 
notes. I would also like to thank the organizers of the XXXIXth Cracow 
Summer School for their invitation and their warm hospitality.

\end{document}